\documentclass[sigconf]{acmart}
\copyrightyear{2022}
\acmYear{2022}
\setcopyright{acmcopyright}
\acmConference[CIKM '22] {Proceedings of the 31st ACM International Conference on Information and Knowledge Management}{October 17--21, 2022}{Atlanta, GA, USA.}
\acmBooktitle{Proceedings of the 31st ACM International Conference on Information and Knowledge Management (CIKM '22), October 17--21, 2022, Atlanta, GA, USA}
\acmPrice{15.00}
\acmISBN{978-1-4503-9236-5/22/10}
\acmDOI{10.1145/3511808.3557354}
\AtBeginDocument{%
  \providecommand\BibTeX{{%
    \normalfont B\kern-0.5em{\scshape i\kern-0.25em b}\kern-0.8em\TeX}}}

\begin{document}

\title{
% 1. Deep Multimodal Recommendation with Context and Interest Enhanced Graph Embeddings\\ 2. Cross-Attention based Context and Interest Enhanced Graph Embeddings for Recommendation\\
% Multi-Interactive Attention based Graph Embedding for Recommendation\\
% Multi-Interactive Attention enhanced  Network for Recommendation
%Graph Embedding enhanced Multi-Interactive Attention Network for Recommendation
HySAGE: A Hybrid Static and Adaptive Graph Embedding Network for Context-Drifting Recommendations
%With Interactive Attention and Graph Embedding % A ... Method for Recommendation\\
% 4. Cross attention graph embedding for Interest enhancement in contextual recommendation\\
% 5. A cross attention based contextual recommendation method with graph embedding
}

\author{Sichun Luo}
\affiliation{%
\department{Department of Computer Science}
  \institution{City University of Hong Kong}
  % \streetaddress{P.O. Box 1212}
  \city{Hong Kong}
  \country{China}
}
\affiliation{%
% \department{Department of Computer Science}
  \institution{City University of Hong Kong Shenzhen Research Institute}
  % \streetaddress{P.O. Box 1212}
  \city{Shenzhen}
  \country{China}
}
\email{sichun.luo@my.cityu.edu.hk}

\author{Xinyi Zhang}
\affiliation{%
\department{Department of Accounting}
  \institution{Capital University of Economics and Business}
  % \streetaddress{P.O. Box 1212}
  \city{Beijing}
  \country{China}
}
\email{zhangxinyi@cueb.edu.cn}

		% China	

\author{Yuanzhang Xiao}
\affiliation{%
\department{Department of Electrical \& Computer Engineering}
  \institution{University of Hawaii at Manoa}
  % \streetaddress{P.O. Box 1212}
  \city{Honolulu}
  \state{HI}
  \country{USA}
}
\email{yxiao8@hawaii.edu}

\author{Linqi Song}
\affiliation{%
\department{Department of Computer Science}
  \institution{City University of Hong Kong}
  % \streetaddress{P.O. Box 1212}
  \city{Hong Kong}
  \country{China}
}
\affiliation{%
% \department{Department of Computer Science}
  \institution{City University of Hong Kong Shenzhen Research Institute}
  % \streetaddress{P.O. Box 1212}
  \city{Shenzhen}
  \country{China}
}
\email{linqi.song@cityu.edu.hk}
\authornote{Corresponding author}

\begin{abstract}

The recent popularity of edge devices and Artificial Intelligent of Things (AIoT) has driven a new wave of contextual recommendations, such as location based Point of Interest (PoI) recommendations and computing resource-aware mobile app recommendations. In many such recommendation scenarios, contexts are drifting over time. For example, in a mobile game recommendation, contextual features like locations, battery, and storage levels of mobile devices are frequently drifting over time. However, most existing graph-based collaborative filtering methods are designed under the assumption of static features. Therefore, they would require frequent retraining and/or yield graphical models burgeoning in sizes, impeding their suitability for context-drifting recommendations. 

In this work, we propose a specifically tailor-made Hybrid Static and Adaptive Graph Embedding (HySAGE) network for context-drifting recommendations. Our key idea is to disentangle the relatively static user-item interaction and rapidly drifting contextual features. Specifically, our proposed HySAGE network learns a relatively static graph embedding from user-item interaction and an adaptive embedding from drifting contextual features. These embeddings are incorporated into an interest network to generate the user interest in some certain context. We adopt an interactive attention module to learn the interactions among static graph embeddings, adaptive contextual embeddings, and user interest, helping to achieve a better final representation. Extensive experiments on real-world datasets demonstrate that HySAGE significantly improves the performance of the existing state-of-the-art recommendation algorithms.
\end{abstract}

\settopmatter{printfolios=true}

\begin{CCSXML}
<ccs2012>
   %     <concept_id>10002951.10003317.10003331.10003271</concept_id>
   %     <concept_desc>Information systems~Personalization</concept_desc>
   %     <concept_significance>500</concept_significance>
   %     </concept>
   <concept>
       <concept_id>10002951.10003317.10003347.10003350</concept_id>
       <concept_desc>Information systems~Recommender systems</concept_desc>
       <concept_significance>500</concept_significance>
       </concept>
 </ccs2012>
\end{CCSXML}

% \ccsdesc[500]{Information systems~Personalization}
\ccsdesc[500]{Information systems~Recommender systems}

\keywords{Recommender system; Context-aware recommendation; Graph embedding; Attention}

\maketitle

\section{Introduction}

With the proliferation of mobile edge devices, contextual information is explored dramatically to show its powerful impact in recommender systems, especially towards personalized recommendations. Compared to traditional context-aware recommendations with commonly used contextual information of time, location, companion, and environmental situation, these recent contextual recommendations focus more on features from the mobile and edge devices. One example is to recommend Point of Interest (PoI) based on location, weather, and social behavior~\cite{aliannejadi2018personalized}. As another example, mobile app recommendations can adapt to the mobile device's resources and usage levels, such as computing power, communication capacity, battery levels, etc~\cite{aliannejadi2021context}. 

In these emerging contextual recommendations, we observe that contexts are often drifting rapidly, compared to the relatively stable user-item interactions. For example, in a mobile game recommendation scenario, contextual features like locations, battery, and storage levels of mobile devices are frequently changing over time and there are quite a few versions of the game tailor-made for different mobile resources; while the user's rating behavior over the mobile game is sparse and relatively static. Recent graph-based deep leaning techniques have been widely used in recommendations. However, some of these graph-based methods~\cite{chen2020enhancing, chen2019joint,tao2020mgat} focused on the user-item interaction and failed to fully exploit the contextual information (or even neglected the contextual information). Other works did not take into account the drifting of contextual information and may cause prohibitively high computation complexity or sparsity problems (e.g., the number of nodes being in the order of user/item-context concatenation pairs) in graph based solutions when the space of drifting contextual information grows large~\cite{wei2019mmgcn,vasile2016meta}. Therefore, these existing works may not be suitable for computing resource constrained mobile devices and edge devices in the Artificial Intelligent of Things (AIoT) systems.

To tackle these problems, we propose a tailor-made graph based solution, termed a Hybrid Static and Adaptive Graph Embedding (\textbf{HySAGE}) network, for context-drifting recommendations. Intuitively, we adopt a hybrid  structure to learn different embeddings for relatively static user-item interaction and rapidly drifting contextual information. By decoupling the adaptive representation of contextual information and the static representation of user-item interaction, our proposed method is especially suitable for context-drifting recommendations. In this way, the drifted contextual attributes would pass through the embedding layer via interactive attention mechanisms and there is no need to re-train the whole graph. Therefore, using such a hybrid structure could potentially save computation resources and retraining time.

Specifically, the proposed HySAGE network first uses user and item relations (e.g., the rating matrix) to construct a bipartite graph and obtain user and item similarity graphs. After that, we adopt a co-occurrence and random walk-based graph embedding algorithm to exploit user and item collaborative embeddings respectively. Meanwhile, multimodal contextual information is also incorporated via various embedding techniques, like specific pre-trained models for texts and images (e.g., Sentence-BERT and ResNet), pre-processing techniques (e.g., normalization and feature crossing) for other categorical and dense contextual features. To reduce the feature dimensionality and learn the feature interaction, the generated feature vectors are fed into the feature crossing layer to learn the higher-order non-linear feature interactions. After obtaining the extracted graph embedding and contextual embedding, we adopt the self-attention mechanism to model the user interests. Instead of the average pooling, the attention mechanism can learn the importance of each component and assign different weights to the components accordingly. As a result, we are able to fuse all the representations from different sources and acquire the final representation for users and items. Finally, these representations are fed into a multi-layer perception to predict the final ratings. Our experiments over four real-world datasets show that our proposed HySAGE outperforms benchmark solutions by up to $20\%$--$30\%$ through effectively processing the drifting contextual information.
%To the best of our knowledge, we are the first to introduce the representation learning from graph embedding, contextual representation learning, and attention-based user interests modeling for dynamic recommendation.

To summarize, the main contributions of this work are three-fold:

$\bullet$ To effectively learn the fast changing contextual information and relative static user-item interaction, we propose a novel end-to-end HySAGE network for context-drifting recommendations, in which the graph embedding module is combined with contextual feature extraction module and user interest mining module to generate a comprehensive representation from different sources.

$\bullet$ We incorporate recent advanced techniques to better learn the comprehensive representation: a co-occurrence and random walk-based graph embedding technique to extract  both global statistical information and local graph information to obtain user and item embeddings accordingly; a multimodal processing technique for jointly exploring multimodal contextual information and other categorical and dense features; a self-attention mechanism to learn the user interest from both the graph embedding and the contextual information embedding; and an interactive attention mechanism to combine different representations into a comprehensive representation.

$\bullet$ We carry out extensive experiments on four real-world datasets to demonstrate the effectiveness of our method (up to $20\%$-$30\%$ gains over benchmarks) and its importance in incorporating contextual information and user interest.
% \sout{We conduct experiments on four real-world datasets, demonstrating the effectiveness of our method. Moreover, we carry out extensive experiments to show the impact of different factors, including the impact of the multimodal incorporation, the effect of the fusion method, etc..}

%This paper is organized as follows. In Section~\ref{sec:related}, we briefly review some related work. We  describe our framework and algorithms in detail in Section~\ref{sec:model}. In Section~\ref{sec:expe}, we conduct some experiments to evaluate the proposed model and go through a thoughtful discussion. Finally, we conclude our work in Section~\ref{sec:conclu}.

\section{Related Work}
\label{sec:related}

% Though CF shows its strength in handling collaborative information, it still suffers from the so-called data sparsity and cold-start problems.

\subsection{Context-Aware Recommendations}
Context-aware recommendation is a popular research direction for two decades, since it utilizes contextual information, such as time, location, social relationship data, and environmental data, to facilitate the recommendation and ease the data sparsity and cold-start problems \cite{xin2019cfm}. Contextual recommendations span from early works with feature engineering techniques to extract categorical and dense features, like movies or news recommendations \cite{adomavicius2011context,li2010contextual}, to recent trends of location and social behavior based PoI recommendations and resource-aware mobile app recommendations~\cite{li2019context,aliannejadi2018personalized,aliannejadi2021context}.

Feature interaction and user interest modeling are two commonly used methods to incorporate contexts in context-aware recommendations. Factorization machines (FM) and its deep learning versions, such as DeepFM, convolutional FM (CFM) and deep cross network (DCN), are able to capture the interactions between different input features and embed them into a low-dimensional latent space~\cite{rendle2010factorization,xin2019cfm,guo2017deepfm,wang2017deep,chen2020efficient}.  Interest modeling, e.g., deep interest network (DIN), deep interest evolution network (DIEN), and deep session interest network (DSIN), enables the incorporation of various contextual features and adopts an attention mechanism to model users' interests based on these features and user-item interactive behaviors~\cite{zhou2018deep,zhou2019deep,feng2019deep}. 

However, these existing methods did not make specific design to deal with the rapid context-drifting problem. Hence, in this work, we will highlight the importance of disentangling relative static user-item interactions and more dynamic contexts, and show the effectiveness of such a tailor-made solution.

\subsection{Graph-Based Recommendations} 
% Collaborative filtering (CF) and content-based recommendation are the conventional ways to make recommendation.
% Neighborhood models are the pioneer in the research of recommender systems. 
% Reference \cite{herlocker1999algorithmic} proposes a user-based collaborative filtering model for personalized rating predictions. 
% Reference \cite{linden2003amazon} further proposes item-based CF models which have a better performance on e-commerce recommendation scenarios.
% With the popularity of deep learning in recent years, CF is incorporated with deep neural networks to yield better performance.
% Reference \cite{he2017neural} firstly proposes NCF (Neural Collaborative Filtering), which replaces the inner product function with a multi-layer tower-shape deep structure as the prediction functions. 
% Similar to \cite{he2017neural}, Reference \cite{chen2019joint} adds embedding transformation layers above user embedding layers and item embedding layers for deep features modeling, and then feeds user and item features into a multi-layer neural network. 
% In \cite{zhang2017joint}, an MLP is placed over the element-wise product of user embedding and item embedding in order to extract user-item interaction patterns. Reference \cite{du2019modeling} further models user-item interactions using a deep convolutional neural network and finally yields a predictive score from feature maps. 
Recently, graph-based models have attracted more attention in recommendations to extract higher-order relations between users and items due to its powerful ability to capture multi-hop relations in the graph. A neural graph collaborative filtering (NGCF) method employs a 3-hop graph neural network to learn user and item embeddings from the user-item bipartite graph~\cite{wang2019neural}.
LightGCN \cite{he2020lightgcn} further improve it by removing the feature transformation and nonlinear activation operation, which contribute little to the model performance.
PinSage~\cite{ying2018graph} utilized a 2-hop graph convolutional neural network and random walk to facilitate the representation learning. 
MMGCN \cite{wei2019mmgcn} proposes a graph convolutional network to learn representations from the multimodal information in multimedia recommendations. A random walk based graph embedding method~\cite{chen2020enhancing} is used to extract high-order collaborative signals from the user-item bipartite graph. Graph neural networks (GNNs) are also introduced to model user's local and global interest for recommendations \cite{wu2019session,xu2019graph,yu2020tagnn}.

Nonetheless, incorporating a large amount of drifting contextual information into the graph based models would lead to an exploding number of nodes (user/item-context concatenation) in the graph, posing difficulties in learning and retraining. In light of this, a specifically designed solution is needed to disentangle user-item interactions and contexts and a more comprehensive design is needed to combine these embeddings learned from different information sources.

\section{Preliminaries}
\label{sec:preliminaries}

We consider a dynamic context-drifting recommender system that consists of $N$ users and $M$ items, and denote the sets of users and items as $\mathcal{U} = \{u_1,...,u_N\}$ and $\mathcal{I} = \{i_1,...,i_M\}$, respectively. Different from
problem formulation of static recommendations, we consider a time horizon of $T$ time slots.

\noindent\textbf{Attributes.} 
At each time slot $t \in \{1,\ldots,T\}$, the attribute of each user $u_n \in \mathcal{U}$ is denoted by a vector $\mathbf{a}_{u_n}(t) \in \mathbb{R}^{d_U}$, and the attribute of each item $i_m \in \mathcal{I}$ by a vector $\mathbf{a}_{i_m}(t) \in \mathbb{R}^{d_I}$. In this way, we model the drifting contexts associated with users and/or items: user attributes (e.g., locations, battery levels of the device) and item attributes (e.g., freshness of content, item descriptions).
\iffalse
We denote the set of all possible user attributes by $\mathcal{A}_U \subset \mathbb{R}^{d_U}$ and the set of all possible user attributes by $\mathcal{A}_I \subset \mathbb{R}^{d_I}$. At each time slot $t \in \{1,\ldots,T\}$, the attribute of each user $u_n \in \mathcal{U}$ is denoted by a vector $\mathbf{a}_{u_n}(t) \in \mathcal{A}_U$, and the attribute of each item $i_m \in \mathcal{I}$ by a vector $\mathbf{a}_{i_m}(t) \in \mathcal{A}_I$. In this way, we model the drifting user attributes (e.g., locations, battery levels of the device) and item attributes (e.g., item descriptions).
\fi

\noindent\textbf{User-Item Interactions.}
Over the course of $T$ time slots, a user $u_n$ may have interaction with item $i_m$ at time $t$ and generate multimodal information (e.g., texts, images, and/or videos from reviewing the item). We denote the contextual information by a vector $\mathbf{c}_{u_n i_m}(t) \in \mathbb{R}^{d_C}$. At the end of $T$ time slots, we use the matrix $\mathbf{Y} \in \{0,1\}^{N \times M}$ to summarize the user-item interactions, where $\mathbf{Y}_{nm}=1$ if user $u_n$ has interaction with item $i_m$ is observed and $\mathbf{Y}_{nm}=0$ otherwise. Note that this interaction matrix can be relaxed by allowing $Y_{nm}\in\mathbb{R}$ to reflect multiple interactive behaviors (like the frequency of listening to some music).

\noindent\textbf{User Interests.}
Each user $u_n$ has time-varying interests in each item $i_m$, denoted by a vector $\mathbf{\theta}_{u_ni_m}(t) \in \mathbb{R}^{d_I}$. 

\noindent\textbf{Ratings.}
We wish to learn the users' ratings of the items at time $t$ based on all the information available up to time $t$. We write the collection of attributes, contextual information, and user interests up to time $t$ by $\mathbf{a}_{u_n}[1:t], \mathbf{a}_{i_m}[1:t]$, $\mathbf{c}_{u_n i_m}[1:t]$, and $\mathbf{\theta}_{u_ni_m}[1:t]$, respectively. Formally, we model user $u_n$'s rating of item $i_m$ at time $t$ as follows
\begin{eqnarray}\label{eqn:rating_model}
R_t\left( \mathbf{e}_{u_n}, \mathbf{e}_{i_m}, \mathbf{a}_{u_n}(t), \mathbf{a}_{i_m}(t), \mathbf{c}_{u_n i_m}[1:t], \mathbf{\theta}_{u_ni_m}(t) \right),
\end{eqnarray}
where $\mathbf{e}_{u_n}$ and $\mathbf{e}_{i_m}$ are vectors representing user $u_n$ and item $i_m$. Note that we use vectors $\mathbf{e}_{u_n}$ and $\mathbf{e}_{i_m}$, instead of indices $n$ and $m$, to represent the user and the item, because more useful information can be encoded into the vectors.

\iffalse
since the user/item attribute/interaction may drift with time.
The sparse rating matrix $Y \in \mathbb{R}^{N \times L}$ records the interactions of the users and items, where $Y_{nl}=1$ if an interaction between user $u_n$ and item $i_l$ is observed and $Y_{nl}=0$ otherwise.
Furthermore, there are a set of $J$ fields of user attributes $\mathcal{A} = \{A_1, ..., A_j\}$ and a set of $K$ fields of item attributes $\mathcal{B} = \{B_1, ..., B_k\}$, while each user and item is associated with a list of attributes $A_u \subset A$ and $B_i \subset B$.
Beyond user-item interactions, we also consider the multimodal features, such as textual description, displaying image, and so on. For simplicity, we denote the modality features as $e_m \in \mathbb{R}^{d_m} $ , where $d_m$ denotes dimension of the features, $m \in M$ is the modality, and $\mathcal{M}$ is the set of modalities.
In this paper, we only consider visual and textual modalities, denoted by $\mathcal{M} = \{v, t\}$, while our method could further extend to other modalities.
Given a user $u$, the task of multimedia recommendation is to recommend a list of items that the user $u$ would be interested in based on the previous information.
\fi

Based on our model \eqref{eqn:rating_model}, we need the following key components:

\textit{(i) Static embeddings of user and item identities $\mathbf{e}_{u_n}$ and $\mathbf{e}_{i_m}$}. This is done by a graph embedding algorithm that captures user-item interactions as well as user and item similarities. Note that the graph embedding produces static embeddings.

\textit{(ii) Adaptive embeddings of time-varying user and item attributes $\mathbf{a}_{u_n}(t)$ and $\mathbf{a}_{i_m}(t)$, and contextual information of user-item interactions $\mathbf{c}_{u_n i_m}(t)$.} The attributes and contextual information can include multimodal information such as numbers, texts, and images. Therefore, we propose a contextual information extraction module to fusion multimodal information into vectors. Note that the embeddings are time-varying, capturing drifting attributes and contextual information.

\textit{(iii) Estimation of time-varying user interests $\mathbf{\theta}_{u_n}(t)$.} Based on user and item attributes $\mathbf{a}_{u_n}[1:t]$ and $\mathbf{a}_{i_m}[1:t]$, and contextual information of user-item interactions $\mathbf{c}_{u_n i_m}[1:t]$, we estimate the user interests $\mathbf{\theta}_{u_ni_m}(t)$.

\textit{(iv) Estimation of user ratings $R_t$ in \eqref{eqn:rating_model}.} We estimate user ratings based on all available information.

To the best of our knowledge, the proposed framework is the only end-to-end method that combines all four components. The resulting recommendation algorithm is context-aware and interest-aware, which boosts the performance. Moreover, it reduces the computational complexity by decoupling the static graph embeddings and the adaptive embeddings of attributes, context, and user interests. In this way, it avoids repeated training of graph embeddings, while still capturing all the information available.

% Models are trained to predict the interacted value $y_{ui}$ in $\textbf{Y}$, given user $u$ and item $i$. Models are expected to learn user preferences and precisely predict a $y_{ui}$ to rank items.

\section{The Proposed Framework}
\label{sec:model}
%\subsection{Overview of The Proposed Framework}

We present HySAGE, a Hybrid Static and Adaptive Graph Embedding for dynamic recommendation.
The framework consists of four major components: a graph embedding module, a contextual information extraction module, a user interest modeling module, and an interactive attention module.

\begin{figure*}[htbp!]
    \centering
    \includegraphics[width=5.6in]{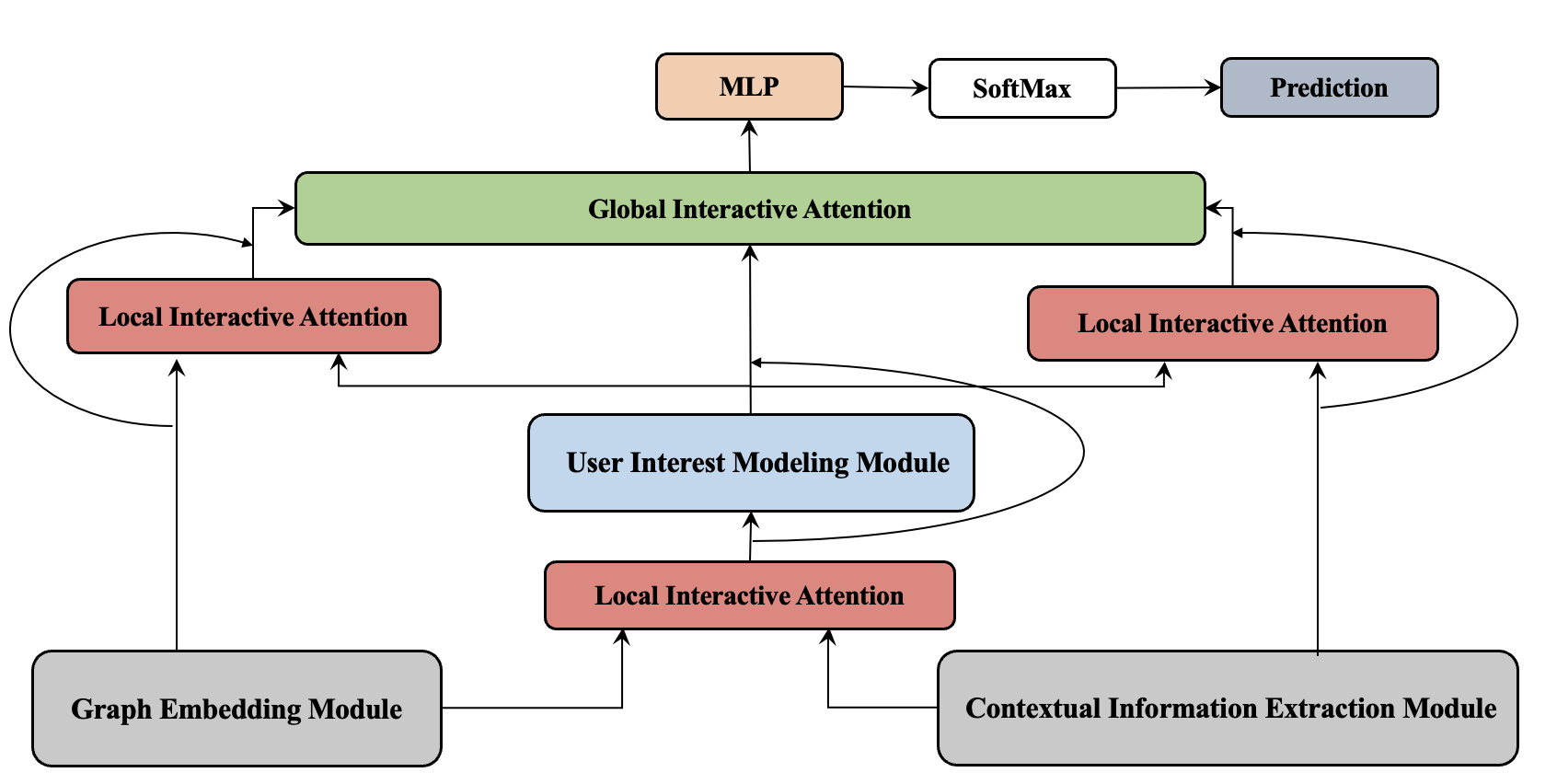}
    \caption{The overall architecture of the proposed model. We first learn the graph embeddings for users and items by co-occurrence and random walk-based techniques.
    Based on the learned graph embeddings, we introduce an attention mechanism to learn the users' interests.
Besides, we extract features from contextual information by pretrained models and use a feature crossing network and an attention layer to learn the hidden representation of contextual information. All of these are concatenated, go through a global interactive attention layer, and are fed to a MLP to get the final recommendation.}
    \label{overall}
\end{figure*}

HySAGE operates in four major steps to perform the recommendation. First, we learn static graph embeddings of users and items by building the user and item bipartite network and mining their high-order collaborative embeddings. 
Second, we obtain adaptive embeddings of contextual information about user-item interaction through (1) representing drifting user and item attributes, (2) adopting pre-trained neural networks to extract multimodal user-item interaction information (e.g., audio, image, text), and (3) using a feature crossing layer to fusion user and item attributes and user-item interaction information, compress the dimension, reduce redundancy and learn high-level feature interactions.
% We transform the sparse rating matrix to dense and low dimension user and item embeddings in this process.
Third, we use the attention mechanism to model the users' recent interests.
Fourth, we use local interactive attention mechanisms to extract bilateral interactions among static graph embeddings, adaptive embeddings of user-item interactions, and adaptive user interests, and use a global interactive attention mechanism to learn the final representations from individual embeddings and their bilateral interactions.
Finally, we train a multi-layer perception (MLP) to predict the users' ratings of the items.
Fig. ~\ref{overall} illustrates our proposed framework.

\subsection{Graph Embedding}
\label{gm1}

\begin{figure}[htbp!]
    \centering
    \includegraphics[width=0.48\textwidth]{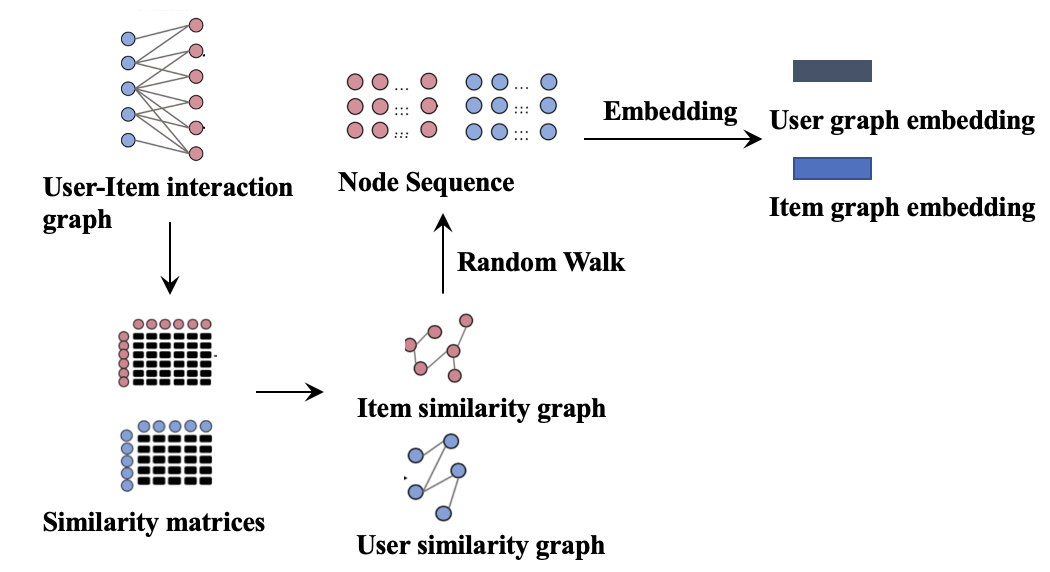}
    \caption{Graph embedding module.}
\label{fig:ge}
\end{figure}

The graph embedding module builds the user and item similarity graphs and learns static graph embeddings $\mathbf{e}_{u_n}$ and $\mathbf{e}_{i_m}$ for each user $u_n \in \mathcal{U}$ and each item $i_m \in \mathcal{I}$. 
We adopt graph embedding method similar to \cite{chen2020enhancing}. Fig.~\ref{fig:ge} illustrates the module.

\textbf{Building User/Item Similarity Graphs.}
We first calculate user and item similarity matrices solely based on the interactions $\mathbf{Y}$ between users and items. 
The user (item) similarity matrix is a $N \times N$ ($M \times M$) matrix, with each element being a co-interaction value between two users (items). 
A widely-used definition of the co-interaction value between two users is the number of items they both interacted with, and that between two items is the number of users who interacted with both of them. Mathematically, the user similarity matrix can be calculated as
\begin{equation}
        \mathbf{S}^{U} = \mathbf{Y} \cdot \mathbf{Y}^{\top} \in \mathbb{R}^{N \times N},
\end{equation}
and the item similarity matrix can be calculated as
\begin{equation}
        \mathbf{S}^{I} = \mathbf{Y}^{\top} \cdot \mathbf{Y} \in \mathbb{R}^{M \times M}.
\end{equation}
Note that our proposed framework can also use other definitions of co-interaction values \cite{sarwar2001item}, such as Pearson correlation, cosine distance, and Jaccard similarity. 

The user similarity matrix defines a user similarity graph $\mathcal{G}_U$, with the set of the nodes as the user set $\mathcal{U}$. There exists an edge between user $u_n$ and user $u_l$ if their co-interaction value $s_{nl}^U$ (i.e., the $(n,l)$-th element of the user similarity matrix $\mathbf{S}^U$) is non-zero (i.e., the two users have interacted with the same item). We define the weights of the edges as the corresponding co-interaction values. The item similarity graph $\mathcal{G}_I$ can be defined in the same way.

\iffalse
The co-interaction value between two user is the number of items they co-interacted with, and co-interaction value between two item is the number of users co-consumed them. $\textbf{Y}$ is the user-item interaction matrix, let $\textbf{M}^{u}$ denotes the user-user co-interaction matrix, or say user-user similarity matrix. The co-interaction value between two user $u$,$i$ can be computed as:
\begin{equation}
    \textbf{M}^{u}_{ui} = \textbf{y}_u * \textbf{y}_i^{\top},
\end{equation}
where $\textbf{y}_u$ and $\textbf{y}_i$ are rows of $\textbf{Y}$. Also, we can compute item co-interaction value in the same way. So, we can compute a user similarity matrix and an item similarity matrix by multiplying the user-item interaction matrix whit its transpose using matrix calculations:

\begin{equation}
        \textbf{M}^{u} = \textbf{Y}*\textbf{Y}^{\top},\textbf{M}^{i} = \textbf{Y}^{\top}*\textbf{Y}.
\end{equation}
Finally we get two similarity matrices -- item similarity matrix $\textbf{M}^{i}$ and user similarity matrix $\textbf{M}^{u}$ and these two matrices can be viewed as the adjacent matrices of two graphs: a user similarity graph $\textbf{G}^u$ and a item similarity graph $\textbf{G}^i$. These two graph brings useful collaborative information that is extracted from u-i interactions.
\fi

\textbf{Mining Collaborative Information.}
% \label{sec:ci}
Based on user and item similarity graphs, we mine collaborative information using graph embedding techniques. We adopt deep learning based graph embedding techniques \cite{grover2016node2vec, perozzi2014deepwalk}, which have been widely applied in network and graph analysis for their ability in extracting node and edge features \cite{cai2018comprehensive}.

To capture the structures of a graph, most graph embedding techniques simulate a random walk process to create node sequences. We describe the process to generate user sequences based on the user similarity graph; the item sequences are generated in the same way. Denote the user sequence of a fixed length $K$ by $\left\{ u_{n_1}, \ldots, u_{n_k}, \ldots, u_{n_K} \right\}$. Given the $k$th node $u_{n_k}$, the next node $u_{n_{k+1}}$ is randomly selected from node $u_{n_k}$'s neighbors according to the following probabilities \cite{grover2016node2vec}:
\begin{equation}
    P\left(u_{n_{k+1}} | u_{n_k}\right) \propto \left\{\begin{array}{ll}
s_{n_k n_{k+1}}^U / p, & \text{if}~u_{n_{k+1}} = u_{n_{k-1}}
\\
s_{n_k n_{k+1}}^U, & 
\text{if}~u_{n_{k+1}}~\text{is a neighbor of}~u_{n_{k-1}}
\\
s_{n_k n_{k+1}}^U / q, & \text{otherwise}
\end{array}\right.,
    \label{eq:tranprob}
\end{equation}
with $p, q > 0$. In other words, the next node is chosen with a probability proportional to the similarity to the current node (measured by the co-interaction value $s_{n_k n_{k+1}}^U$), moderated by parameters $p,q$. For example, we can choose $p>1$ to discourage the selection of the previous node $u_{n_{k-1}}$ and $q<1$ to encourage the selection of nodes that are not neighbors of the previous node $u_{n_{k-1}}$. 
As we can see from \eqref{eq:tranprob}, a user more similar to the current node, as measured by a higher co-interaction value, is more likely to be selected as the next node in the sequence. Hence, this sampling process helps us to better capture collaborative information.

%It is proved that nodes distributions in the node sequences generated by random walks will have a very similar distribution as word distribution in documents \cite{cai2018comprehensive} in natural language processing (NLP). 
% So researchers usually adopt some NLP model to process these node sequences.

\iffalse
We then adopt a random walk process to generate input sequences: starting from a node in the graph, we choose a neighboring node as the next node according to neighboring nodes' transition probability $P(c_{i+1} | c_{i})$, where $c_{i}$ is the current node; $c_{i+1}$ is the next node. In order to avoid being trapped in a local area, we use weighted random walk, where the transition probability depends on the distance between the previous node and the next node and the weight of edges $w_{c_i,c_{i+1}}$ connecting $c_{i}$ and $c_{i+1}$, i.e.,  
\begin{equation}
    P(c_{i+1} | c_{i}) \propto a_{pq}(c_{i-1},c_{i+1}) \times w_{c_i,c_{i+1}},
    \label{eq:tranprob}
\end{equation}
where $c_{i-1}$ is the previous node and $a_{pq}(c_{i-1},c_{i+1})$ is a control parameter
\begin{equation}
    a_{pq}(c_{i-1},c_{i+1}) = \left\{\begin{matrix}
\frac{1}{p}, \text{distance}(c_{i-1},c_{i+1}) = 0,
\\1, \text{distance}(c_{i-1},c_{i+1}) = 1,
\\\frac{1}{q}, \text{distance}(c_{i-1},c_{i+1}) = 2,
\end{matrix}\right.
\label{eq:jump}
\end{equation}
where $p$ controls the probability of revisiting previous nodes and $q$ controls the probability of visiting distance nodes.
Starting from every node in the graph, we generate multiple fixed length walk sequences. Using such a random walk strategy, similar users/items tend to appear in the same walk sequence and thus sharing similar context nodes, which will help to better extract collaborative information.
\fi

Given all the $K$-node sequences, we use the global co-occurrence based method in \cite{pennington2014glove} to learn the user embeddings $\mathbf{e}_{u_n}$. The co-occurrence between two users $u_{n}$ and $u_{l}$, denoted $O_{nl}$, is defined as the empirical frequency of these two users appearing in the same user sequence. We use user embeddings to estimate the co-occurrence according to
\begin{equation}
    \widehat{O}_{nl} = \text{exp}\left(\mathbf{e}_{u_n}^\top \tilde{\mathbf{e}}_{u_l} + b_{u_n} + \tilde{b}_{u_l}\right), 
\end{equation}
where $b_{u_n} \in \mathbb{R}$ is the bias, $\tilde{\mathbf{e}}_{u_l}$ and $\tilde{b}_{u_l}$ are the embedding and the bias of nodes $u_l$ when node $u_l$ appears second in a two-node pair. Note that the embedding and the bias of node $u_l$ when it appears first in a two-node pair are ${\mathbf{e}}_{u_l}$ and ${b}_{u_l}$.

We aim to learn the user embeddings $\mathbf{e}_{u_n}$ and $\tilde{\mathbf{e}}_{u_n}$ that minimize the discrepancy between the estimated co-occurrence $\widehat{O}_{nl}$ and the actual co-occurrence $O_{nl}$. We define the loss function of the training process as
\begin{eqnarray}
    & & \mathcal{L}^{GE}(\mathbf{e}_{u_1}, \ldots, \mathbf{e}_{u_N}, b_{u_1}, \ldots, b_{u_N}) \\
    &=& \sum_{(u_n,u_l)} \left( \log\widehat{O}_{nl} - \log O_{nl} \right)^2 \nonumber \\
    &=& \sum_{(u_n,u_l)} \left( \mathbf{e}_{u_n}^\top \tilde{\mathbf{e}}_{u_l} + b_{u_n} + \tilde{b}_{u_l} - \log O_{nl} \right)^2. \nonumber  
\end{eqnarray}

The training process can be performed by a two-layer neural network whose input is one-hot embedding of the user nodes. After the training, we set $\mathbf{e}_{u_n} \leftarrow \mathbf{e}_{u_n} + \tilde{\mathbf{e}}_{u_n}$.

Note that although we use static graph embeddings of the users and items, our framework can update the graph embeddings through graph convolution \cite{kipf2016semi}.

\subsection{Contextual Information Extraction}
\begin{figure}[htbp!]
    \centering
    \includegraphics[width=0.47\textwidth]{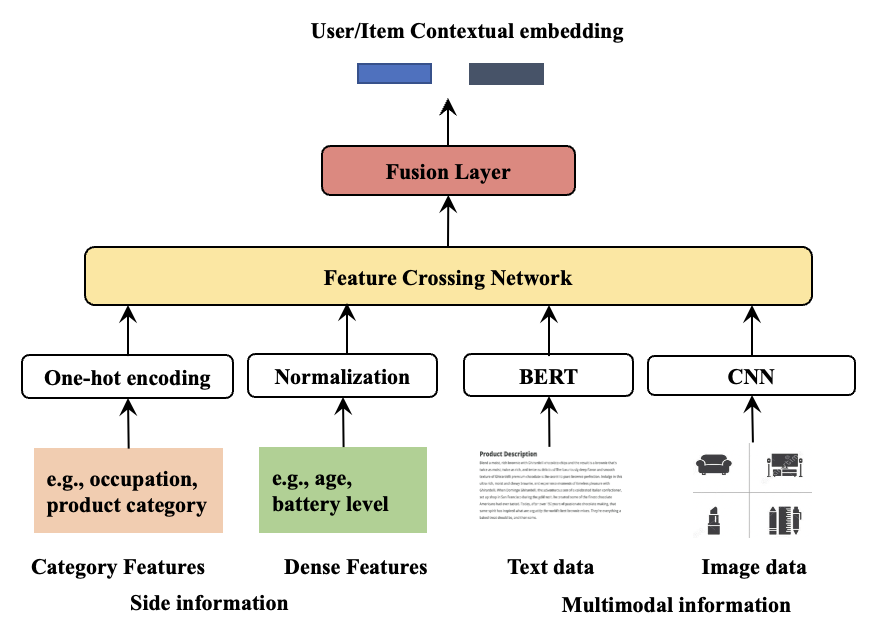}
    \caption{Contextual information extraction module.}
\label{fig:context}
\end{figure}

The contextual information extraction module learns the adaptive embeddings of drifting user and item attributes $\mathbf{a}_{u_n}(t), \mathbf{a}_{i_m}(t)$ and user-item interactions $\mathbf{c}_{u_n i_m}(t)$. These adaptive embeddings are obtained from multimodal information (e.g., categorical, numerical, textual, image). Based on the embeddings, the module uses a feature crossing layer and the attention mechanism for better representation learning of the feature interaction. Fig.~\ref{fig:context} illustrates the contextual information extraction module.

{\textbf{Category Features.}} 
Some of the user and item attributes are categorical data (e.g., occupations of the users, product categories of the items). We use one-hot embedding of the categorical data initially, and feed the sparse one-hot vectors to the feature crossing layer to obtain dense vectors later.

{\textbf{Dense Features.}} 
Dense features (e.g., user ages) are normalized using min-max scalers and then fed into the feature crossing layer.

{\textbf{Text Data.}} Text data (e.g., titles and description of items, review by the users) contains critical information. In this work, we use a pretrained model called Sentence-BERT \cite{reimers2019sentence} to process the raw data and get fixed-length vectors as output.
%However, text data is unstructured and hence cannot be directly input into the system. Therefore, some language models like Word2vec, Doc2vec, BERT, and LSTM could be used to process the textual modal.
% Before applying the pretrained model, we get the feature embedding of the textual modal.

{\textbf{Image Data.}}  Image data (e.g., display images of items, pictures uploaded by the users) contains crucial hidden information. Due to the high dimensionality of raw images, here we adopt a pre-trained convolutional neural network (CNN), namely ResNet\cite{he2016deep}, to process the images and get fixed-length vectors as output.

\iffalse
Like text data, the image data is too large and noisy to input into the model directly. 
Instead of the whole image, we need the feature extracted from the image. Therefore, we adopt a pre-trained model to process the original image and get the feature vector . 
More precisely, a convolutional neural network (CNN) is adopted to extract the image. As the mainstream and popular model in the computer vision domain, CNN has lots of open-source pre-trained models like VGG, Resnet, and Alexnet trained on ImageNet. We use it for extracting image features and comparing their performance in detail in the latter part. For the whole process, first, we preprocess the image by resizing and normalizing it. Then we send it to a pre-trained CNN model and get a fixed-length feature vector as the output. The image feature could be fused with other modals for better representation learning.
\fi

%We collect the one-hot embedding of the user (item) categorical features and preprocessed user (item) dense features into the embedding of user (item) attributes $\mathbf{a}_{u_n}(t)$ ($\mathbf{a}_{i_m}(t)$).

After one-hot embedding of categorial features, preprocessing of dense features, and embedding of text and image data, we obtain embeddings of user and item attributes $\mathbf{a}_{u_n}(t), \mathbf{a}_{i_m}(t)$ and user-item interactions $\mathbf{c}_{u_n i_m}(t)$. We write the embedding vector of all the features as
\begin{eqnarray}
    \mathbf{x}_0 = \text{Concat}\left(\mathbf{a}_{u_n}(t), \mathbf{a}_{i_m}(t), \mathbf{c}_{u_n i_m}(t)\right)
    \in \mathbb{R}^{d_U + d_I + d_C}
\end{eqnarray}
where $\text{Concat}()$ is the concatenation operation.
Instead of using the embedding vector $\mathbf{x}_0(t)$ directly, we pass it through a feature crossing network to learn high-order interaction across features \cite{wang2017deep} and a fusion layer to reduce the dimensionality.

{\textbf{The Feature Crossing Network.}}  
The feature crossing network consists of $L$ feature crossing layers. The output $\mathbf{x}_{l+1} \in \mathbb{R}^{d_U + d_I + d_C}$ of layer $l+1$ is obtained by \cite{wang2017deep}
\begin{equation}
    \mathbf{x}_{l+1} = \mathbf{x}_0 \mathbf{x}_l^\top \mathbf{w}_l + \mathbf{b}_l + \mathbf{x}_l,
\end{equation}
where $\mathbf{x}_{l} \in \mathbb{R}^{d_U + d_I + d_C}$ is the output of layer $l$, $\mathbf{w}_l, \mathbf{b}_l \in \mathbb{R}^{d_U + d_I + d_C}$ are weight and bias parameters. The feature crossing is achieved by the product $\mathbf{x}_0 \mathbf{x}_l^\top$. A deeper feature crossing network captures higher-order interaction across features.

{\textbf{The Fusion Layer.}} 
% At this step, an intuitively idea is to simply sum up the weighted contextual feature vectors to represent the user/item. 
% However, a simple weighted average could introduce too much noisy information since these information may have redundancy.
Given the embedding vector $\mathbf{x}_0$ and the high-order cross feature embedding $\mathbf{x}_L$, the fusion layer uses the self-attention mechanism to obtain the final embedding of the contextual information. As opposed to simply concatenating or adding $\mathbf{x}_0$ and $\mathbf{x}_L$, the self-attention mechanism assigns different weights to features and hence can focus on more important features.
The self-attention mechanism can be mathematically expressed as:
\begin{equation}
    \mathbf{e}_c = \sum_{i\in\{0,L\}} \frac{\text{exp}(\text{tanh}(\mathbf{w}_i^\top \cdot \mathbf{x}_i + b_i))}{\sum_{i^\prime\in\{0,L\}} \text{exp}(\text{tanh}(\mathbf{w}_{i^\prime}^\top \cdot \mathbf{x}_{i^\prime} + b_{i^\prime})) } \mathbf{x}_i,
\end{equation}
where $\mathbf{w}_i$ and $b_i$ are the weights and the bias of the attention layer, and $\mathbf{e}_c$ is the final output for contextual information extraction module.

\iffalse
\begin{equation}
    \mathbf{e}_c = \sum_{i=1}^n \frac{\text{exp}(\text{tanh}(R_i \cdot W_i + b_i))}{\sum_{i^\prime=1}^n \text{exp}(\text{tanh}(R_{i^\prime} \cdot W_{i^\prime} + b_{i^\prime})) } R_i,
\end{equation}
\fi

% Here, we choose to stack further neural transformations to extract higher-order features. To accommodate with the noise and irrelevant aspects extracted above, we employ a attention mechanism to further abstract higher-level features as follows: 
% In the last part, we have achieved multimodal embeddings. 
% Commonly used fusion method include feature level fusion and decision level fusion.
% However, the decision level fusion is relatively hard to implement since multimodal embeddings are not sufficient to make the decision.
% We would use the feature level fusion, or called early fusion.
% By fusing the multimodal feature at an early stage, we could better learn the feature interaction.
% An intuitive way is to concatenate all multimodal embeddings and send them forward to MLP together with all other embeddings. This method, though simple, is not efficient and appropriate to get the final prediction. Because multimodal embeddings may have crossing features. This would cause redundancy and thus not optimal. Also, some part of the embedding is less critical.
% Therefore, we adopt the self-attention mechanism to get the embedding with lower dimensions and could focus on the more important part.

\subsection{User Interest Modeling}
Previous works show user interests are diverse and dynamic, and may have a considerable impact on the recommendation performance \cite{zhou2019deep, zhou2018deep}.
Therefore, we utilize the user-item interactions to learn the user interests.

To learn user $u_n$'s interest in a target item $i_m$, we randomly select $K$ items that this user has interacted with. We select $K$ random items, instead of the most recent $K$ items, because we do not assume the availability of time and sequence information. Denote this set of $K$ items as $\mathcal{I}_{u_n} = \left\{ i_{u_n,1}, i_{u_n,2}, \ldots, i_{u_n,K} \right\}$, where $K$ is a hyperparameter.
In Section~4.1, we have obtained the embedding $\mathbf{e}_{i_m}$ of the target item $i_m$ and the embeddings $\mathbf{e}_{i_{u_n,k}}$ of the selected items $i_{u_n,k}$ for $k=1,\ldots,K$. We learn user $u_n$'s ``valuation'' (relative to the target item $i_m$) of the $k$ selected item $i_{u_n,k}$ through a MLP:
\begin{equation}\label{eqn:valuation}
v_{k}=  \text{MLP}\left(\mathbf{e}_{i_{u_n,k}},~\mathbf{e}_{i_m},~\mathbf{e}_{i_{u_n,k}}-\mathbf{e}_{i_m}\right) \in \mathbb{R}.
\end{equation}
Then we calculate user $u_n$'s interest in the target item $i_m$ by passing the embeddings of the selected items through a self-attention layer:
\begin{equation}\label{eqn:interest}
\begin{split}
    \mathbf{\theta}_{u_n i_m} = 
    \sum_{k=1}^K \frac{\text{exp}(u_k)}{\sum^K_{j=1} \text{exp}(u_j)} \cdot \mathbf{e}_{i_{u_n,k}}.
\end{split}
\end{equation}

\begin{figure}[htbp!]
    \centering
    \includegraphics[width=0.28\textwidth,height=0.3\textwidth]{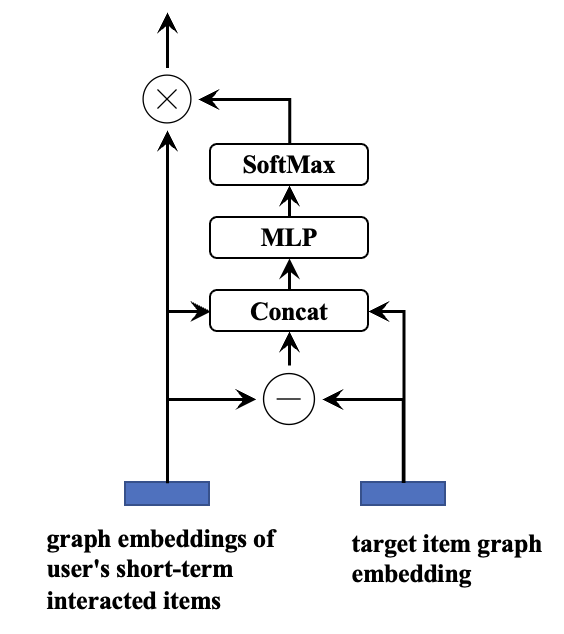}
    % \caption{We concatenate the graph embedding of the target item and user's recent interacted item, feed them into an MLP, and then use the softmax to get the attention weight. We multiply the attention weight with the user's recent interacted item embeddings and use the weighted sum to represent user interests.}
\caption{
    User Interest Modeling}
\label{fig:selfatt}
\end{figure}

The user interest module is illustrated in Fig.~\ref{fig:selfatt}.

\subsection{Multi-Interactive Learning}

At this point, we have obtained a set of latent vectors for the users and items in interaction-based feature learning, denoted as ${V_u, V_l, V_c}$, representing user interest vector, contextual vector, and collaborative vector respectively.
Previous works indicate low-order and high-order interaction could contribute to model performance.
Here we highlight the importance of learning high-order interactive information from different feature representations. The intuition is that different representations describe different aspects of the users and items. Therefore, directly concatenate them is not enough for well representation learning.
Therefore, we propose a Multi interactive learning module. More specifically, we learn the importance of each component by an attention mechanism. Besides, we explicitly learn the representation interaction by a feature crossing structure.

% user interest u
% context l
% collobarative c

\textbf{Local Interaction Learning.}
We regard the contextual representation of the user and item as their inherent attrib{ut}es, thus the latent contextual vector learned above describes the characteristics of users and items.
Besides, the collaborative feature provides rich information abo{ut} the user's general preferences.
Moreover, the user interest model the personalized preference with respect to the specific items.
Hence, each latent vector expresses the user/item in a different way and a single latent vector does not represent the users and items to their fullness. 
Therefore, we derive a learning process for latent representation of the interaction between interest features, collaborative features, and contextual features. Specifically, we learn the interaction between every two components.
Similar to the user interest part, we introd{uc}e an attention mechanism to quantify the importance of representation interaction and calculate the scaled attention representation as follows.
Let $V$ represent a set of feature vectors from collaborative vector, contextual vector, and interest vector respectively. $V_t$ is a subset of $V$.
The $R_{cl}$ represents the feature interaction between the candidate user interest and contextual information.
Firstly, we simply concatenate the contextual information with the candidate user interest representation to get the interactive feature representation, formulated by:
\begin{equation}
V_{ut}^k = \text{Concat}(V_u^k, V_t).
\end{equation}

Then we adopt the attention mechanism to emphasize different parts of user interest by assigning different weights. The formula for the final representation $R_{ut}$ is
\begin{equation}
R_{ut} = \sum^K_{k=1} \frac{\text{exp}(\text{tanh}(V_{ut}^k \cdot W_{ut}^k + b_{ut}^k)}{\sum^K_{k^\prime=1} \text{exp}(\text{tanh}(V_{ut}^{k^\prime} \cdot W_{ut}^{k^\prime} + b_{ut}^{k^\prime}))} V_{ut}^k.
\end{equation}

Moreover, let the $R_{uc}$ represents the feature interaction between the candidate user interest and collaborative information.
Similarly, we get the final representation formula for $R_{uc}$.

\begin{equation}
\begin{split}
& V_{uc}^k = \text{Concat}(V_u^k, V_c)\\
& R_{uc}^k = \sum^K_{k=1} \frac{\text{exp}(\text{tanh}(V_{uc}^k \cdot W_{uc}^k + b_{uc}^k)}{\sum^K_{k^\prime=1} \text{exp}(\text{tanh}(V_{uc}^{k^\prime} \cdot W_{uc}^{k^\prime} + b_{uc}^{k^\prime}))} V_{uc}^k.
\end{split}
\end{equation}

\textbf{Global Interaction Learning.}
In the global interaction learning part, we still adopt attention, which is a technique that is often used in natural language processing applications \cite{vaswani2017attention}.
Specifically, we concatenate all the embedding, denoted as $R_{ai} = [R_1, R_2, ... R_n]$ where $R_{ai}$ in $\{R_1, R_2, ... R_n\}$ is a kind of embedding extracted in the previous module. Then we use the attention mechanism to get the final output for the global interaction learning module$R_{ao}$ , and the formula  is denoted as:
\begin{equation}
    R_{ao} = \sum_{i=1}^n \frac{\text{exp}(\text{tanh}(R_i \cdot W_i + b_i))}{\sum_{i^\prime=1}^n \text{exp}(\text{tanh}(R_{i^\prime} \cdot W_{i^\prime} + b_{i^\prime})) } R_i,
\end{equation}
% where 

% \begin{equation}
% \begin{split}
% & MultiHead(x) = Concat(head_1, head_2, ... , head_h)W^O \\
% & head_i = Attention(l_b W_i^Q, l_b W^K_i, l_b W_i^V).
% \end{split}
% \end{equation}
% where $head_i \in \{V_i,V_l,V_c,V_{lc},V_{ic},V_{il}\}$.
% By learning the global interaction of different module, our proposed method could attentively assign attention score for each module. This adaptive process would enable the system maintain a satisfactory performance even when some module lack of enough information.

\subsection{DNN Prediction Module}
A tower-shaped MLP structure is often used for predicting ratings in recommender systems \cite{fan2019graph, he2017neural}. By utilizing neural structures, the prediction functions have strong ability in handing non-linear relations and can capture user-item interactions far better than simple inner product function. 
However, these existing works did not consider using graph embedding and user interest to obtain initial features.

In our framework, we adopt a tower-shaped MLP as the prediction network, as shown at the top of Fig.~\ref{overall}.
The input of the MLP includes the global feature representation generated by the global interaction module.
To make accurate predictions, we also sample 4 un-interacted items for a user to generate negative samples. 
The MLP has fully connected layers, uses softmax as the activation function of the output layer, and uses rectified linear unit (ReLU) as the activation function of the other layers. 
We use a mini-batch gradient descent and Adaptive Moment Estimation (Adam) optimization algorithm to train the neural network.
\iffalse
\begin{equation}
     \min \sum_{(u,i) \in (\mathcal{P} \cup \mathcal{N}) }  r_{ui}\log\hat{r}_{ui} + (1 - r_{ui}) \log(1 - \hat{r}_{ui}),
\end{equation}
where $\mathcal{P}$ and $\mathcal{N}$ are the sets of positive and negative samples, $\hat{r}_{ui}$ and $r_{ui}$ are the predicted and actual ratings.
\fi

\section{Experiments}
\label{sec:expe}

In this section, we perform experiments on real-world datasets from different domains for performance evaluation. We also analyze the components for our proposed framework. We aim to answer these following research questions:

\textbf{RQ1}: How does our model performs in comparison to baseline methods? Will mining and encoding collaborative information help? If the method can outperform others, why using graph-enhanced method can help?  

\textbf{RQ2}: How does the incorporation of user interest and contextual information will affect the recommendation performance?

\textbf{RQ3}: How do hyperparameters (i.e., the embedding size, the random walk length) affect the performance?

\subsection{Experiment Setup}
\subsubsection{Datasets}
We perform experiments on four real-world datasets to evaluate our proposed HySAGE framework. 
Among the four datasets, \emph{Kaggle-Movie} and \emph{MovieLens-100K} are movie datasets, and \emph{Yelp} and \emph{Amazon-Kindle} are review datasets.
Table~\ref{dataset} summarizes the statistics of the datasets. Detailed description of the datasets are as follows.

\textbf{$\bullet$ Yelp:} This is from Yelp-2018 challenge. The task is to recommend business webpages to users. 
To ensure the quality of dataset, we remove the users who give less than five reviews and the businesses who receive less than five reviews.

\textbf{$\bullet$ Amazon-Kindle:} This is from the Amazon review data. The task is to recommend e-books to users.
We use the 10-core setting to ensure the quality of dataset. 
% The final dataset contains 14355 users, 15884 items, and 367478 records. 
Compared to the Yelp dataset, it has similar numbers of users and items, but much fewer interactions. Therefore, it allows us to test our method when user-item interactions are sparse.

\textbf{$\bullet$ MovieLens-100K:} This is a canonical movie recommendation dataset widely used for evaluating recommendation algorithms.

\textbf{$\bullet$ Kaggle-Movie:} This is an extended MovieLens dataset released on Kaggle. We remove the movies with less than two records.

% Table ~\ref{dataset} summarizes the detailed statistics about the four aforementioned datasets after preprocessing. 

%\subsection{Dataset Description}

%We made experiments on four public dataset to evaluate performances:

%\textbf{- Yelp:} Yelp Dataset is taken from Yelp-2018 challenge. .To ensure the quality of dataset, we sample the dataset to get a sub-dataset we make sure that every user gives at least 5 ratings and every business at least receives 5 ratings from users.

%\textbf{- Amazion-Kindle:} A dataset collected from Amazon review data collections. we apply a 10-core setting to ensure the quality of dataset. 

% Finally we get a dataset contains 14355 users and 15884 items, in total 367478 records. It has similar user number and item number to the yelp-2018 dataset, however, the interactions are sparser than the yelp-dataset, we want to test if our method still works when user-item interactions are sparse.

%\textbf{- MovieLens-100K:} A stable movie rating dataset collected by GroupLens. This dataset is widely used for evaluating recommendation algorithms.

%\textbf{- Kaggle-MovieDataset:} The movie rating dataset is released on Kaggle. This dataset is an extended edition of movielens dataset.We remove those movies which have less than 2 records.

% This dataset has a similar number of ratings to MovieLens 100K, but interactions are sparser.

%Table \ref{dataset} give statistics of these dataset. We transformed them into implicit datasets.

%\subsection{Performance Comparison}
\subsubsection{Baselines}
We compare our proposed HySAGE algorithm with the following canonical and state-of-the-art algorithms.

\textbf{$\bullet$ Item popularity:} We simply rank items by its popularity and utilize no collaborative information.

\textbf{$\bullet$ Bayesian Personalized Ranking (BPR) \cite{rendle2009bpr}:}  Instead of optimizing over point-wise user-item ratings, this model is trained to rank interacted items higher than those with no interaction.

\textbf{$\bullet$ MF \cite{koren2009matrix} + NS:} a traditional matrix factorization method with negative sampling to enhance the data. 

\textbf{$\bullet$ Neural Collaborative Filtering (NCF) \cite{he2017neural}:} a deep learning-based collaborative filtering method using deep tower-shaped MLPs.

\textbf{$\bullet$ JRL \cite{zhang2017joint}:} a multi-layer neural network incorporating element-wise products of the user and item embedding.

\textbf{$\bullet$ NeuMF \cite{he2017neural}:} a deep learning based collaborative filtering model combining generalized matrix factorization \cite{he2017neural} and NCF. 

\textbf{$\bullet$ GEM-RS \cite{chen2020enhancing}:} GEM-RS uses graph embedding to learn collaborative information from user-item interaction matrix. 
% GEM-RS performs well in many tasks and is our main benchmark.

\textbf{$\bullet$ PinSage \cite{ying2018graph}:} a large scale graph convolutional network that aggregates random walks and graph convolution operations.

\textbf{$\bullet$ Neural Graph Collaborative Filtering (NGCF) \cite{wang2019neural}:} a state-of-the-art neural graph collaborative filter method that mines collaborative signals from the graph structure.
% a representative and state-of-the-art
% GCN model for recommendation. 

\begin{table}[t]
\centering
\caption{Dataset description}
\label{dataset}

\scalebox{0.86}[0.86]{
\begin{tabular}{c c c c c}
\hline
Dataset   & \# of user & \# of item & \# of rating & sparsity \\ \hline
Yelp & 21,284      & 16,771      & 984,000      & 99.72\%  \\ 
Amazion-Kindle & 14,355      & 15,884      & 367,478      & 99.84\%  \\ 
MovieLens-100K & 943      & 1,682      & 100,000      & 94.12\%  \\ 
Kaggle-Movie & 670      & 5,977      & 96,761      & 98.55\%  \\ \hline

\end{tabular}
}

\end{table}

\subsection{Experiments Settings}
For movie datasets, we use the pre-trained ResNet50 model \cite{he2016deep} to extract the visual features from the movie posters, and use the Sentence-BERT \cite{reimers2019sentence} to extract the textual features from the movie descriptions.
For all deep-learning-based methods, the activation functions are ReLU, the learning rates are 0.001 to 0.01, the $L_2$ regulations are $10^{-7}$, and the batch sizes are 256. 
For all four datasets, we set the embedding size to 64 for fair comparison. %The node number of the first layer is twice as the embedding size. 
In NCF, NeuMF, and GEM-RS, the number of nodes is halved by each layer and the last layer has 8 nodes. In JRL, there are three layers with the same number of nodes. 
In HySAGE, the walk length is 20 and the walk number is 100. 
We perform a grid search on the hyperparameters (e.g., learning rate, dropout rate) and select the best for each method according to the performance
on the validation set.

We adopt the commonly used leave-one-out evaluation strategy, which reserves the latest interacted item of each user as the test item. Then we generate a list of items and rank them using the predicted scores. We use two metrics to evaluate the performance: the Hit Ratio (HR), which measures the chance that our recommendation list contains users' interested items, and a weighted version of HR, termed Normalized Discounted Cumulative Gain (NDCG), which puts more weights on items that are ranked higher in the recommendation list.
All performance metrics are computed on the test set and averaged over five runs.

\begin{table*}[h]
\centering
\caption{Performance Comparison}
\label{performance}
\scalebox{0.93}[0.93]{
\begin{tabular}{c cc cc cc cc }
\hline
\multirow{2}{*}{Model} & \multicolumn{2}{c}{Yelp}         & \multicolumn{2}{c}{Amazon-Kindle} & \multicolumn{2}{c}{MovieLens-100K} & \multicolumn{2}{c}{Kaggle-Movie} \\
                  & HR@10           & NDCG@10         & HR@10            & NDCG@10         & HR@10            & NDCG@10          & HR@10           & NDCG@10         \\  \hline
Item popularity   & 0.2940          & 0.1611          & 0.2755           & 0.1466          & 0.4189           & 0.2337           & 0.4829          & 0.2713          \\  \hline
MF+NS             & 0.6712          & 0.4011          & 0.5925           & 0.3691          & 0.6351           & 0.3549           & 0.5956          & 0.3691          \\ 
BPR               & 0.7239          & 0.4212          & 0.6925           & 0.4176          & 0.5762           & 0.3021           & 0.5887          & 0.3753          \\ \hline
JRL               & 0.6889          & 0.4217          & 0.6834           & 0.4331          & 0.6473           & 0.3657           & 0.6349          & 0.3881          \\  
NCF               & 0.7245          & 0.4316          & 0.7134           & 0.4372          & 0.6525           & 0.3789           & 0.6617          & 0.3973          \\ 
NeuMF             & 0.7688          & 0.4734          & 0.7194           & 0.4469          & 0.6522           & 0.3918           & 0.6647          & 0.4036          \\  \hline
GEM-RS            & 0.7886 & {0.4920} & {0.7637}  & {0.4911} & {0.6681}  & {0.3950}  & {0.6885} & {0.4280} \\ 
PinSage            & 0.7625 & {0.4371} & {0.7184}  & {0.4424} & {0.6660}  & {0.4010}  & {0.6706} & {0.3986} \\ 
NGCF            & 0.7833 & {0.4616} & {0.7207}  & {0.4544} & {0.6706}  & {0.4229}  & {0.6734} & {0.4031} \\ \hline
HySAGE            & \textbf{0.9527} & \textbf{0.8571} & \textbf{0.9529}  & \textbf{0.8095} & \textbf{0.8348}  & \textbf{0.6832}  & \textbf{0.8841} & \textbf{0.7443} \\
\hline
Improvement & 
\textbf{20.81\%} & \textbf{74.21\%} & \textbf{24.77\%} & \textbf{64.83\%} & \textbf{24.95\%} & \textbf{72.96\%} & \textbf{28.41\%} & \textbf{73.90\%} \\ \hline %\cline{1-1}
		
% Improvement over NCF            & \textbf{8.85\%} & \textbf{13.99\%} & \textbf{7.10\%}  & \textbf{12.33\%} & \textbf{2.40\%}  & \textbf{4.25\%}  & \textbf{4.05\%} & \textbf{7.73\%} \\ \hline
% Improvement over NeuMF            & \textbf{2.58\%} & \textbf{3.93\%} & \textbf{6.16\%}  & \textbf{9.84\%} & \textbf{2.44\%}  & \textbf{0.82\%}  & \textbf{3.58\%} & \textbf{6.05\%} \\ \hline
\end{tabular}
}
\end{table*}

\subsection{Performance Comparison \textbf{(RQ1)}}
Table \ref{performance} reports the overall performance of all methods.
The best results and the improvements over the second best are highlighted in bold. 
First, deep learning based algorithms (JRL, NCF, NeuMF, GEM-RS) generally perform better than traditional methods (Item popularity, MF+NS, BPR). 
The item-popularity method is much worse than the other methods, which stresses the importance of mining collaborative information between users and items.

Second, neural based methods and graph based methods have better performance over the other methods. These methods can learn the nonlinear representation better with graph structures and neural networks.
For example, NCF uses neural networks to learn collaborative information.
This shows that it is promising to introduce neural networks and graph structures into the system for enhanced representation learning and interaction modeling.

Third, our proposed HySAGE approach consistently achieves best scores on all four datasets.
%We can observe that the improvements gained by HySAGE are consistent and stable.
% outperforms the state-of-the-art  GEM-RS algorithm. 
% This is due to the fact that our proposed HySAGE can mine explicit and implicit relations between items and users in the embedding feature pretraining phase using graph embedding method.
On average, the relative improvement of HySAGE against the best baseline is 24.74\% for HR@10 and 71.48\% for NDCG@10.
For Yelp, the sparsest dataset, it gains a 20.81\% improvement compared to the best baseline.
Even for Kindle, the dataset without user feature data, it gains a 24.77\% improvement over the best baseline.
This result shows the effectiveness of HySAGE on datasets with different characteristics. Moreover, the significant performance gap between HySAGE and GEM-RS validates that user interest mining and context-aware user-item representation learning in HySAGE capture more knowledge about users' diverse rating behaviors on items by considering both the individual characteristics of users and items and their interactions.

% We could obeserve the importance of 

\subsection{Ablation Study \textbf{(RQ2)}}
To prove the effectiveness of each critical module in our proposed framework, we designed a set of ablation experiments. 
Specifically, we repeat the experiment by removing one module from the proposed HySAGE model and test the performance of these incomplete models on four datasets. The incomplete models are as follow:

\textbf{• HySAGE-Variation 1 (w/o multimodal information)}: The proposed HySAGE without the feature extraction of multimodal information in the contextual information extraction module.

\textbf{• HySAGE-Variation 2 (w/o side information)}: The proposed HySAGE without the feature extraction of side information in the contextual information extraction module.

\textbf{• HySAGE-Variation 3 (w/o user interest)}: The proposed HySAGE without the user interest mining module.

\textbf{• HySAGE-Variation 4 (w/o multi-interactive)}: The proposed HySAGE without the multi-interactive learning module.

From Table \ref{ablation}, we observe that the HySAGE achieves the best results on all datasets, which verifies the importance of each critical module.
Among them, the experimental results of HySAGE w/o
user interest mining drop sharply, which proves that the lack of attentively learning the user interest could significantly decrease the learning ability of the framework.
Without the side information extraction or multimodal information extraction, the performance of the framework still decreases considerably. 
This reflects that the contextual information contains abundant auxiliary information, which improves the model performance. 
However, compared with the other modules, the impact of
contextual information is relatively small.
The contextual information module is like a residual network to supplement the extra content information to the HySAGE framework.
Throughout these three ablation experiments, it turns out
that each module improves the model performance from
different aspects and is meaningful.

\begin{table*}[]
\centering
\caption{Ablation study of recommendation performance in two datasets}
\label{ablation}
\scalebox{0.8}[0.8]{
\begin{tabular}{l cccccc cccccc}
\hline
\multirow{2}{*}{Model} & \multicolumn{6}{c}{Movielens}  & \multicolumn{6}{c}{Kaggle} \\ 
 &  HR@5   & NDCG@5 & HR@10  & NDCG@10 & HR@20  & NDCG@20 & HR@5   & NDCG@5 & HR@10  & NDCG@10 & HR@20  & NDCG@20 \\ \hline
HySAGE                 & 0.8147 & 0.7348 & 0.8838 & 0.7572  & 0.9417 & 0.7718  & 0.8349 & 0.7471 & 0.9038 & 0.7694  & 0.9519 & 0.7817  \\
HySAGE-Variation1      & 0.7335 & 0.6301 & 0.8332 & 0.6619  & 0.9103 & 0.7086  & 0.7754 & 0.7517 & 0.8991 & 0.7697  & 0.9488 & 0.7826  \\ 
HySAGE-Variation2      & 0.7489 & 0.6553 & 0.8348 & 0.6832  & 0.9104 & 0.7023  & 0.8096 & 0.7210 & 0.8841 & 0.7443  & 0.9411 & 0.7584  \\ 
HySAGE-Variation3      & 0.5047 & 0.3549 & 0.6702 & 0.4076  & 0.8213 & 0.4445  & 0.5406 & 0.3831 & 0.6792 & 0.4256  & 0.8133 & 0.4590  \\ 
HySAGE-Variation4  &0.7385&0.6329&0.8255&0.6611&0.9077&0.6846&0.7914&0.7056&0.8763&0.7328&0.9359&0.7472\\ \hline
\end{tabular}
}
\end{table*}

% \begin{table*}[h]
% \centering
% \caption{Ablation study of recommendation performance in two datasets}
% \label{ablation}
% \scalebox{0.8}[0.8]{
% \begin{tabular}{l|ll|ll|ll|ll|ll|ll}
% \hline
%  \toprule
% \multirow{}{}{Algorithm}             & \multicolumn{6}{|l|}{Movielens}                         & \multicolumn{6}{|l}{Kaggle}                            \\
%               \cmidrule(l){2-13} 
%               & HR@5   & NDCG@5 & HR@10  & NDCG@10 & HR@20  & NDCG@20 & HR@5   & NDCG@5 & HR@10  & NDCG@10 & HR@20  & NDCG@20 \\
%               \hline
% HySAGE          & 0.8147 & 0.7348 & 0.8838 & 0.7572  & 0.9417 & 0.7718  & 0.8349 & 0.7471 & 0.9038 & 0.7694  & 0.9519 & 0.7817  \\
% HySAGE-Variation 1   & 0.7335 & 0.6301 & 0.8332 & 0.6619  & 0.9103 & 0.7086  & 0.7754 & 0.7517 & 0.8991 & 0.7697  & 0.9488 & 0.7826  \\
% HySAGE-Variation 2 & 0.7489 & 0.6553 & 0.8348 & 0.6832  & 0.9104 & 0.7023  & 0.8096 & 0.7210 & 0.8841 & 0.7443  & 0.9411 & 0.7584  \\
% HySAGE-Variation 3     & 0.5047 & 0.3549 & 0.6702 & 0.4076  & 0.8213 & 0.4445  & 0.5406 & 0.3831 & 0.6792 & 0.4256  & 0.8133 & 0.4590 \\
% \hline
% \end{tabular}}
% \end{table*}

\subsection{Analysis of HySAGE \textbf{(RQ3)}}
{\luo{\subsubsection{Hyper-parameter Studies}}}
We study the impact of different model settings of HySAGE.

\begin{figure}[bp!]
    \centering
    \includegraphics[width=0.45\textwidth]{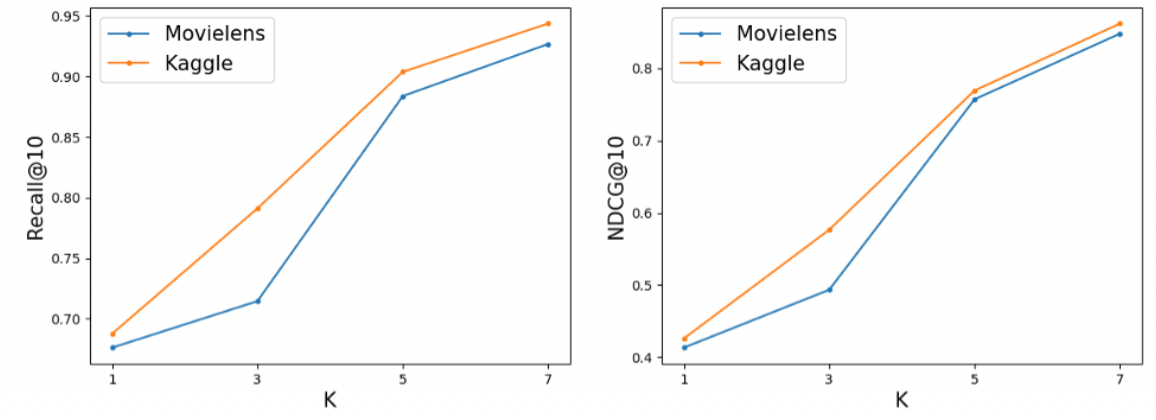}
%   \hspace{0.01in}
%   \includegraphics[width=0.23\textwidth]{images/p1.jpg}
    % \includegraphics[width=0.25\textwidth]{}
    % \includegraphics[width=0.25\textwidth]{images/p1.jpg}
    \caption{Performance of HySAGE w.r.t. different
parameter $K$ on Movielens and Kaggle.}
\label{fig:k}
\end{figure}

\textbf{Impact of the Hyperparameter $K$.}
In the user interest mining module, we randomly sample $K$ items and derive user potential interest using the attention mechanism.
We study the impact of the hyperparameter $K$, which determines the model capacity. 
% Table~\ref{hyperK} 
Figure~\ref{fig:k} lists the model performance of HySAGE under different $K$. 
We can see that the performance of the HySAGE improves with the increase of $K$.
In essence, higher $K$ allows HySAGE to learn user interest information from more data, and therefore achieve better representation learning of user interest.

\textbf{Impact of the Embedding Size.}
We study how the embedding size (the dimension of the embedding/latent features) affects the performance of HySAGE. 
Table~\ref{table:embedding} shows the performance when the embedding size varies from $32$ to $128$. For the two datasets tested, the embedding size of $32$ achieves the best performance. As we increase the embedding size, there is an degrade in performance. This is because even though high-dimensional feature vectors contain more information, more computation complexity is needed. When the embedding size is too large, the convergence of the training is slower due to the large number of parameters. Larger embedding sizes may also introduce redundant information or overfitting, which deteriorate the performance.

\begin{table}[]
\centering
\caption{Performance under different embedding sizes}
\scalebox{0.8}[0.8]{
\label{table:embedding}
\begin{tabular}{llllll}
\hline
\multicolumn{2}{l}{Embedding size}                       & 32     & 64     & 96     & 128    \\ \hline
\multirow{6}{*}{Movielens} & HR@5    & 0.8354 & 0.8147 & 0.7635 & 0.7532\\
                          & NDCG@5  & 0.7418 & 0.7348 & 0.6601 & 0.6597 \\ \cline{2-6} 
                           & HR@10   & 0.9031 & 0.8838 & 0.8643 & 0.8478 \\ 
                          & NDCG@10 & 0.7638 & 0.7572 & 0.6925 & 0.6903 \\ \cline{2-6} 
                          & HR@20   & 0.9583 & 0.9417 & 0.9279 & 0.9289 \\ 
                           & NDCG@20 & 0.7776 & 0.7718 & 0.7087 & 0.7109 \\ \hline
\multirow{6}{*}{Kaggle}    & HR@5    & 0.8590 & 0.8349 & 0.7917 & 0.7767 \\ 
                           & NDCG@5  & 0.7748 & 0.7471 & 0.6958 & 0.6781 \\ \cline{2-6} 
                           & HR@10   & 0.9171 & 0.9038 & 0.8736 & 0.8589 \\ 
                           & NDCG@10 & 0.7937 & 0.7694 & 0.7204 & 0.7021 \\ \cline{2-6} 
                           & HR@20   & 0.9645 & 0.9519 & 0.9353 & 0.9270 \\ 
                         & NDCG@20 & 0.8055 & 0.7817 & 0.7350 & 0.7186 \\ \hline
\end{tabular}}
\end{table}

\begin{table}[]
\centering
\caption{Performance under different random walk lengths}
\scalebox{0.8}[0.8]{
\label{table:walk}
\begin{tabular}{lllll}
\hline
\multicolumn{2}{l}{Random Walk Length}                   & 5      & 10     & 20     \\ \hline
\multicolumn{1}{l}{\multirow{6}{*}{Movielens}} & HR@5    & 0.6550 & 0.7190 & 0.8147 \\ 
\multicolumn{1}{l}{}                           & NDCG@5  & 0.5337 & 0.6258 & 0.7348 \\ \cline{2-5} 
\multicolumn{1}{l}{}                           & HR@10   & 0.7743 & 0.8148 & 0.8838 \\ 
\multicolumn{1}{l}{}                           & NDCG@10 & 0.5718 & 0.6566 & 0.7572 \\ \cline{2-5} 
\multicolumn{1}{l}{}                           & HR@20   & 0.8778 & 0.8993 & 0.9417 \\ 
\multicolumn{1}{l}{}                           & NDCG@20 & 0.5979 & 0.6780 & 0.7718 \\ \hline
\multicolumn{1}{l}{\multirow{6}{*}{Kaggle}}    & HR@5    & 0.6963 & 0.8036 & 0.8349 \\ 
\multicolumn{1}{l}{}                           & NDCG@5  & 0.5969 & 0.7181 & 0.7471 \\ \cline{2-5} 
\multicolumn{1}{l}{}                           & HR@10   & 0.7919 & 0.8688 & 0.9038 \\ 
\multicolumn{1}{l}{}                           & NDCG@10 & 0.6272 & 0.7393 & 0.7694 \\ \cline{2-5} 
\multicolumn{1}{l}{}                           & HR@20   & 0.8727 & 0.9264 & 0.9519 \\ 
\multicolumn{1}{l}{}                           & NDCG@20 & 0.6473 & 0.7540 & 0.7817 \\ \hline
\end{tabular}}
\end{table}

% \begin{figure}[htbp!]
%     \centering
%     \includegraphics[width=0.45\textwidth]{images/rwf2.png}
%     \caption{HySAGE performance with different random walk length}
% \label{fig:rwf}
% \end{figure}

% \begin{figure}[htbp]
%     \centering
%     \subfigure[SubCaption\_1]{
%         \includegraphics[width=0.2\textwidth]{images/p1.jpg}
%         \label{label_for_cross_ref_1}
%     }
%     \subfigure[SubCaption\_2]{
% 	\includegraphics[width=0.2\textwidth]{images/p1.jpg}
%         \label{label_for_cross_ref_2}
%     }
% %     \quad    %用 \quad 来换行
% %     \subfigure[SubCaption_3]{
% %     	\includegraphics[width=2.5in]{PrintScreen_0003}
% %         \label{label_for_cross_ref_3}
% %     }
% %     \subfigure[SubCaption_4]{
% % 	\includegraphics[width=2.5in]{Photos/PrintScreen_0004}
% %         \label{label_for_cross_ref_4}
% %     }
% %     \caption{This is a Demo of $2\times 2$}
% %     \label{fig.1}
% \end{figure}

\textbf{Impact of the Random Walk Length.} 
We also study how the random walk length affects the performance. The random walk length is the number of steps taken in the random walk. The larger the walk length is, the more likely a walk is to visit further nodes and discover more complex graph structures. We test different walk lengths ranging from 5 to 20. The results are displayed in Table~\ref{table:walk}. 
We can see that the performances of models improve continuously as the walk length increases. Under large walk lengths, similar users and items are more likely to appear in the same walk sequence. Hence, longer walkers produce more expressive and representative embeddings.

\subsection{Further Discussions}

% \textbf{Computational overhead in context drifting settings.} 
For context drifting settings, the traditional GNN model would need to train the graph frequently due to the fast drifting context.
In contrast, our proposed model could save the computational overhead since the graph does not need to be trained continually. The contextual information extraction module would adaptively extract useful information from the drifting context, and the model could still work well.

% \textbf{ discussions.} 

% Note that our framework could also fit into other situations with appropriate revision. For example, in this paper, we discuss the situation that the user-item interaction (i.e., graph structure) is relatively stable. 
% However, when it comes to the situation where the user-item interaction drifts rapidly, then our proposed model would also work by replacing the transductive graph neural networks with inductive neural networks.

\section{Conclusion}
\label{sec:conclu}
In this paper, we have proposed HySAGE, a Context and Interest Enhanced Graph Embedding technique to boost the performance of multimedia recommendations. We build a bipartite graph from user-item interactions and use random walk-based graph embedding techniques to extract user and item embeddings. The graph embedding is incorporated with attention mechanism to mine the user potential interest, then joint with contextual embeddings that are extracted from multimedia and side information to make multimedia recommendations. Experiments on four real datasets demonstrate the effectiveness of our proposed framework and show significant benefits over existing state-of-the-art algorithms.

%Experiments on four public dataset demonstrate that explicitly encode collaborative information in user and item embedding vectors is helpful to recommendation task and noticeably lifts performances. We show that our model outperforms several basline methods, including deep recommendation models and state-of-the-art. After adding contextual information, our method still work well and can reach higher performance in the experiment, this fact indicates that our model has more potential if we combine more multimedia and user information. Besides, we made experiments to study the enhancing effect and hyper-parameters. 

\begin{acks}
This work was supported in part by the Changsha Science and Technology Program International and Regional Science and Technology Cooperation Project under Grants kh2201026, the Hong Kong RGC grant ECS 21212419, the Technological Breakthrough Project of Science, Technology and Innovation Commission of Shenzhen Municipality under Grants JSGG20201102162000001, InnoHK initiative, the Government of the HKSAR, Laboratory for AI-Powered Financial Technologies, the Hong Kong UGC Special Virtual Teaching and Learning (VTL) Grant 6430300, and the Tencent AI Lab Rhino-Bird Gift Fund.
\end{acks}
\bibliographystyle{ACM-Reference-Format}
\balance
\bibliography{sample-base}


\begin{thebibliography}{38}

%%% the Web address.

\ifx \showCODEN    \undefined \def \showCODEN     #1{\unskip}     \fi
\ifx \showDOI      \undefined \def \showDOI       #1{#1}\fi
\ifx \showISBNx    \undefined \def \showISBNx     #1{\unskip}     \fi
\ifx \showISBNxiii \undefined \def \showISBNxiii  #1{\unskip}     \fi
\ifx \showISSN     \undefined \def \showISSN      #1{\unskip}     \fi
\ifx \showLCCN     \undefined \def \showLCCN      #1{\unskip}     \fi
\ifx \shownote     \undefined \def \shownote      #1{#1}          \fi
\ifx \showarticletitle \undefined \def \showarticletitle #1{#1}   \fi
\ifx \showURL      \undefined \def \showURL       {\relax}        \fi
% invisible to TeX
\providecommand\bibfield[2]{#2}
\providecommand\bibinfo[2]{#2}
\providecommand\natexlab[1]{#1}
\providecommand\showeprint[2][]{arXiv:#2}

\bibitem[Adomavicius and Tuzhilin(2011)]%
        {adomavicius2011context}
\bibfield{author}{\bibinfo{person}{Gediminas Adomavicius} {and}
  \bibinfo{person}{Alexander Tuzhilin}.} \bibinfo{year}{2011}\natexlab{}.
\newblock \showarticletitle{Context-aware recommender systems}.
\newblock In \bibinfo{booktitle}{\emph{Recommender systems handbook}}.
  \bibinfo{publisher}{Springer}, \bibinfo{pages}{217--253}.
\newblock


\bibitem[Aliannejadi and Crestani(2018)]%
        {aliannejadi2018personalized}
\bibfield{author}{\bibinfo{person}{Mohammad Aliannejadi} {and}
  \bibinfo{person}{Fabio Crestani}.} \bibinfo{year}{2018}\natexlab{}.
\newblock \showarticletitle{Personalized context-aware point of interest
  recommendation}.
\newblock \bibinfo{journal}{\emph{ACM Transactions on Information Systems
  (TOIS)}} \bibinfo{volume}{36}, \bibinfo{number}{4} (\bibinfo{year}{2018}),
  \bibinfo{pages}{1--28}.
\newblock


\bibitem[Aliannejadi et~al\mbox{.}(2021)]%
        {aliannejadi2021context}
\bibfield{author}{\bibinfo{person}{Mohammad Aliannejadi},
  \bibinfo{person}{Hamed Zamani}, \bibinfo{person}{Fabio Crestani}, {and}
  \bibinfo{person}{W~Bruce Croft}.} \bibinfo{year}{2021}\natexlab{}.
\newblock \showarticletitle{Context-aware Target Apps Selection and
  Recommendation for Enhancing Personal Mobile Assistants}.
\newblock \bibinfo{journal}{\emph{ACM Transactions on Information Systems
  (TOIS)}} \bibinfo{volume}{39}, \bibinfo{number}{3} (\bibinfo{year}{2021}),
  \bibinfo{pages}{1--30}.
\newblock


\bibitem[Cai et~al\mbox{.}(2018)]%
        {cai2018comprehensive}
\bibfield{author}{\bibinfo{person}{Hongyun Cai}, \bibinfo{person}{Vincent~W
  Zheng}, {and} \bibinfo{person}{Kevin Chen-Chuan Chang}.}
  \bibinfo{year}{2018}\natexlab{}.
\newblock \showarticletitle{A comprehensive survey of graph embedding:
  Problems, techniques, and applications}.
\newblock \bibinfo{journal}{\emph{IEEE Transactions on Knowledge and Data
  Engineering}} \bibinfo{volume}{30}, \bibinfo{number}{9}
  (\bibinfo{year}{2018}), \bibinfo{pages}{1616--1637}.
\newblock


\bibitem[Chen et~al\mbox{.}(2020b)]%
        {chen2020efficient}
\bibfield{author}{\bibinfo{person}{Chong Chen}, \bibinfo{person}{Min Zhang},
  \bibinfo{person}{Weizhi Ma}, \bibinfo{person}{Yiqun Liu}, {and}
  \bibinfo{person}{Shaoping Ma}.} \bibinfo{year}{2020}\natexlab{b}.
\newblock \showarticletitle{Efficient non-sampling factorization machines for
  optimal context-aware recommendation}. In
  \bibinfo{booktitle}{\emph{Proceedings of The Web Conference 2020}}.
  \bibinfo{pages}{2400--2410}.
\newblock


\bibitem[Chen et~al\mbox{.}(2019)]%
        {chen2019joint}
\bibfield{author}{\bibinfo{person}{Wanyu Chen}, \bibinfo{person}{Fei Cai},
  \bibinfo{person}{Honghui Chen}, {and} \bibinfo{person}{Maarten de Rijke}.}
  \bibinfo{year}{2019}\natexlab{}.
\newblock \showarticletitle{Joint Neural Collaborative Filtering for
  Recommender Systems}.
\newblock \bibinfo{journal}{\emph{arXiv preprint arXiv:1907.03459}}
  (\bibinfo{year}{2019}).
\newblock


\bibitem[Chen et~al\mbox{.}(2020a)]%
        {chen2020enhancing}
\bibfield{author}{\bibinfo{person}{Weijing Chen}, \bibinfo{person}{Weigang
  Chen}, {and} \bibinfo{person}{Linqi Song}.} \bibinfo{year}{2020}\natexlab{a}.
\newblock \showarticletitle{Enhancing Deep Multimedia Recommendations Using
  Graph Embeddings}. In \bibinfo{booktitle}{\emph{2020 IEEE Conference on
  Multimedia Information Processing and Retrieval (MIPR)}}. IEEE,
  \bibinfo{pages}{161--166}.
\newblock


\bibitem[Fan et~al\mbox{.}(2019)]%
        {fan2019graph}
\bibfield{author}{\bibinfo{person}{Wenqi Fan}, \bibinfo{person}{Yao Ma},
  \bibinfo{person}{Qing Li}, \bibinfo{person}{Yuan He}, \bibinfo{person}{Eric
  Zhao}, \bibinfo{person}{Jiliang Tang}, {and} \bibinfo{person}{Dawei Yin}.}
  \bibinfo{year}{2019}\natexlab{}.
\newblock \showarticletitle{Graph Neural Networks for Social Recommendation}.
\newblock \bibinfo{journal}{\emph{arXiv preprint arXiv:1902.07243}}
  (\bibinfo{year}{2019}).
\newblock


\bibitem[Feng et~al\mbox{.}(2019)]%
        {feng2019deep}
\bibfield{author}{\bibinfo{person}{Yufei Feng}, \bibinfo{person}{Fuyu Lv},
  \bibinfo{person}{Weichen Shen}, \bibinfo{person}{Menghan Wang},
  \bibinfo{person}{Fei Sun}, \bibinfo{person}{Yu Zhu}, {and}
  \bibinfo{person}{Keping Yang}.} \bibinfo{year}{2019}\natexlab{}.
\newblock \showarticletitle{Deep session interest network for click-through
  rate prediction}.
\newblock \bibinfo{journal}{\emph{arXiv preprint arXiv:1905.06482}}
  (\bibinfo{year}{2019}).
\newblock


\bibitem[Grover and Leskovec(2016)]%
        {grover2016node2vec}
\bibfield{author}{\bibinfo{person}{Aditya Grover} {and} \bibinfo{person}{Jure
  Leskovec}.} \bibinfo{year}{2016}\natexlab{}.
\newblock \showarticletitle{node2vec: Scalable feature learning for networks}.
  In \bibinfo{booktitle}{\emph{22nd International Conference on Knowledge
  Discovery and Data Mining}}. \bibinfo{pages}{855--864}.
\newblock


\bibitem[Guo et~al\mbox{.}(2017)]%
        {guo2017deepfm}
\bibfield{author}{\bibinfo{person}{Huifeng Guo}, \bibinfo{person}{Ruiming
  Tang}, \bibinfo{person}{Yunming Ye}, \bibinfo{person}{Zhenguo Li}, {and}
  \bibinfo{person}{Xiuqiang He}.} \bibinfo{year}{2017}\natexlab{}.
\newblock \showarticletitle{DeepFM: a factorization-machine based neural
  network for CTR prediction}.
\newblock \bibinfo{journal}{\emph{arXiv preprint arXiv:1703.04247}}
  (\bibinfo{year}{2017}).
\newblock


\bibitem[He et~al\mbox{.}(2016)]%
        {he2016deep}
\bibfield{author}{\bibinfo{person}{Kaiming He}, \bibinfo{person}{Xiangyu
  Zhang}, \bibinfo{person}{Shaoqing Ren}, {and} \bibinfo{person}{Jian Sun}.}
  \bibinfo{year}{2016}\natexlab{}.
\newblock \showarticletitle{Deep residual learning for image recognition}. In
  \bibinfo{booktitle}{\emph{Proceedings of the IEEE conference on computer
  vision and pattern recognition}}. \bibinfo{pages}{770--778}.
\newblock


\bibitem[He et~al\mbox{.}(2020)]%
        {he2020lightgcn}
\bibfield{author}{\bibinfo{person}{Xiangnan He}, \bibinfo{person}{Kuan Deng},
  \bibinfo{person}{Xiang Wang}, \bibinfo{person}{Yan Li},
  \bibinfo{person}{Yongdong Zhang}, {and} \bibinfo{person}{Meng Wang}.}
  \bibinfo{year}{2020}\natexlab{}.
\newblock \showarticletitle{Lightgcn: Simplifying and powering graph
  convolution network for recommendation}. In
  \bibinfo{booktitle}{\emph{Proceedings of the 43rd International ACM SIGIR
  conference on research and development in Information Retrieval}}.
  \bibinfo{pages}{639--648}.
\newblock


\bibitem[He et~al\mbox{.}(2017)]%
        {he2017neural}
\bibfield{author}{\bibinfo{person}{Xiangnan He}, \bibinfo{person}{Lizi Liao},
  \bibinfo{person}{Hanwang Zhang}, \bibinfo{person}{Liqiang Nie},
  \bibinfo{person}{Xia Hu}, {and} \bibinfo{person}{Tat-Seng Chua}.}
  \bibinfo{year}{2017}\natexlab{}.
\newblock \showarticletitle{Neural collaborative filtering}. In
  \bibinfo{booktitle}{\emph{Proceedings of the 26th international conference on
  world wide web}}. \bibinfo{pages}{173--182}.
\newblock


\bibitem[Kipf and Welling(2016)]%
        {kipf2016semi}
\bibfield{author}{\bibinfo{person}{Thomas~N Kipf} {and} \bibinfo{person}{Max
  Welling}.} \bibinfo{year}{2016}\natexlab{}.
\newblock \showarticletitle{Semi-supervised classification with graph
  convolutional networks}.
\newblock \bibinfo{journal}{\emph{arXiv preprint arXiv:1609.02907}}
  (\bibinfo{year}{2016}).
\newblock


\bibitem[Koren et~al\mbox{.}(2009)]%
        {koren2009matrix}
\bibfield{author}{\bibinfo{person}{Yehuda Koren}, \bibinfo{person}{Robert
  Bell}, {and} \bibinfo{person}{Chris Volinsky}.}
  \bibinfo{year}{2009}\natexlab{}.
\newblock \showarticletitle{Matrix factorization techniques for recommender
  systems}.
\newblock \bibinfo{journal}{\emph{Computer}} \bibinfo{volume}{42},
  \bibinfo{number}{8} (\bibinfo{year}{2009}), \bibinfo{pages}{30--37}.
\newblock


\bibitem[Li et~al\mbox{.}(2010)]%
        {li2010contextual}
\bibfield{author}{\bibinfo{person}{Lihong Li}, \bibinfo{person}{Wei Chu},
  \bibinfo{person}{John Langford}, {and} \bibinfo{person}{Robert~E Schapire}.}
  \bibinfo{year}{2010}\natexlab{}.
\newblock \showarticletitle{A contextual-bandit approach to personalized news
  article recommendation}. In \bibinfo{booktitle}{\emph{Proceedings of the 19th
  international conference on World wide web}}. \bibinfo{pages}{661--670}.
\newblock


\bibitem[Li et~al\mbox{.}(2019)]%
        {li2019context}
\bibfield{author}{\bibinfo{person}{Yang Li}, \bibinfo{person}{Yadan Luo},
  \bibinfo{person}{Zheng Zhang}, \bibinfo{person}{Shazia Sadiq}, {and}
  \bibinfo{person}{Peng Cui}.} \bibinfo{year}{2019}\natexlab{}.
\newblock \showarticletitle{Context-aware attention-based data augmentation for
  poi recommendation}. In \bibinfo{booktitle}{\emph{2019 IEEE 35th
  International Conference on Data Engineering Workshops (ICDEW)}}. IEEE,
  \bibinfo{pages}{177--184}.
\newblock


\bibitem[Pennington et~al\mbox{.}(2014)]%
        {pennington2014glove}
\bibfield{author}{\bibinfo{person}{Jeffrey Pennington},
  \bibinfo{person}{Richard Socher}, {and} \bibinfo{person}{Christopher
  Manning}.} \bibinfo{year}{2014}\natexlab{}.
\newblock \showarticletitle{Glove: Global vectors for word representation}. In
  \bibinfo{booktitle}{\emph{Proceedings of the 2014 Conference on Empirical
  Methods in Natural Language Processing (EMNLP)}}.
  \bibinfo{pages}{1532--1543}.
\newblock


\bibitem[Perozzi et~al\mbox{.}(2014)]%
        {perozzi2014deepwalk}
\bibfield{author}{\bibinfo{person}{Bryan Perozzi}, \bibinfo{person}{Rami
  Al-Rfou}, {and} \bibinfo{person}{Steven Skiena}.}
  \bibinfo{year}{2014}\natexlab{}.
\newblock \showarticletitle{Deepwalk: Online learning of social
  representations}. In \bibinfo{booktitle}{\emph{Proceedings of the 20th ACM
  SIGKDD international conference on Knowledge discovery and data mining}}.
  \bibinfo{pages}{701--710}.
\newblock


\bibitem[Reimers and Gurevych(2019)]%
        {reimers2019sentence}
\bibfield{author}{\bibinfo{person}{Nils Reimers} {and} \bibinfo{person}{Iryna
  Gurevych}.} \bibinfo{year}{2019}\natexlab{}.
\newblock \showarticletitle{Sentence-bert: Sentence embeddings using siamese
  bert-networks}.
\newblock \bibinfo{journal}{\emph{arXiv preprint arXiv:1908.10084}}
  (\bibinfo{year}{2019}).
\newblock


\bibitem[Rendle(2010)]%
        {rendle2010factorization}
\bibfield{author}{\bibinfo{person}{Steffen Rendle}.}
  \bibinfo{year}{2010}\natexlab{}.
\newblock \showarticletitle{Factorization machines}. In
  \bibinfo{booktitle}{\emph{2010 IEEE International conference on data
  mining}}. IEEE, \bibinfo{pages}{995--1000}.
\newblock


\bibitem[Rendle et~al\mbox{.}(2009)]%
        {rendle2009bpr}
\bibfield{author}{\bibinfo{person}{Steffen Rendle}, \bibinfo{person}{Christoph
  Freudenthaler}, \bibinfo{person}{Zeno Gantner}, {and} \bibinfo{person}{Lars
  Schmidt-Thieme}.} \bibinfo{year}{2009}\natexlab{}.
\newblock \showarticletitle{{BPR}: Bayesian personalized ranking from implicit
  feedback}. In \bibinfo{booktitle}{\emph{Proceedings of the 25th Conference on
  Uncertainty in Artificial Intelligence}}. \bibinfo{pages}{452--461}.
\newblock


\bibitem[Sarwar et~al\mbox{.}(2001)]%
        {sarwar2001item}
\bibfield{author}{\bibinfo{person}{Badrul Sarwar}, \bibinfo{person}{George
  Karypis}, \bibinfo{person}{Joseph Konstan}, {and} \bibinfo{person}{John
  Riedl}.} \bibinfo{year}{2001}\natexlab{}.
\newblock \showarticletitle{Item-based collaborative filtering recommendation
  algorithms}. In \bibinfo{booktitle}{\emph{Proceedings of the 10th
  international conference on World Wide Web}}. \bibinfo{pages}{285--295}.
\newblock


\bibitem[Tao et~al\mbox{.}(2020)]%
        {tao2020mgat}
\bibfield{author}{\bibinfo{person}{Zhulin Tao}, \bibinfo{person}{Yinwei Wei},
  \bibinfo{person}{Xiang Wang}, \bibinfo{person}{Xiangnan He},
  \bibinfo{person}{Xianglin Huang}, {and} \bibinfo{person}{Tat-Seng Chua}.}
  \bibinfo{year}{2020}\natexlab{}.
\newblock \showarticletitle{MGAT: multimodal graph attention network for
  recommendation}.
\newblock \bibinfo{journal}{\emph{Information Processing \& Management}}
  \bibinfo{volume}{57}, \bibinfo{number}{5} (\bibinfo{year}{2020}),
  \bibinfo{pages}{102277}.
\newblock


\bibitem[Vasile et~al\mbox{.}(2016)]%
        {vasile2016meta}
\bibfield{author}{\bibinfo{person}{Flavian Vasile}, \bibinfo{person}{Elena
  Smirnova}, {and} \bibinfo{person}{Alexis Conneau}.}
  \bibinfo{year}{2016}\natexlab{}.
\newblock \showarticletitle{Meta-prod2vec: Product embeddings using
  side-information for recommendation}. In
  \bibinfo{booktitle}{\emph{Proceedings of the 10th ACM Conference on
  Recommender Systems}}. ACM, \bibinfo{pages}{225--232}.
\newblock


\bibitem[Vaswani et~al\mbox{.}(2017)]%
        {vaswani2017attention}
\bibfield{author}{\bibinfo{person}{Ashish Vaswani}, \bibinfo{person}{Noam
  Shazeer}, \bibinfo{person}{Niki Parmar}, \bibinfo{person}{Jakob Uszkoreit},
  \bibinfo{person}{Llion Jones}, \bibinfo{person}{Aidan~N Gomez},
  \bibinfo{person}{{\L}ukasz Kaiser}, {and} \bibinfo{person}{Illia
  Polosukhin}.} \bibinfo{year}{2017}\natexlab{}.
\newblock \showarticletitle{Attention is all you need}. In
  \bibinfo{booktitle}{\emph{Advances in neural information processing
  systems}}. \bibinfo{pages}{5998--6008}.
\newblock


\bibitem[Wang et~al\mbox{.}(2017)]%
        {wang2017deep}
\bibfield{author}{\bibinfo{person}{Ruoxi Wang}, \bibinfo{person}{Bin Fu},
  \bibinfo{person}{Gang Fu}, {and} \bibinfo{person}{Mingliang Wang}.}
  \bibinfo{year}{2017}\natexlab{}.
\newblock \showarticletitle{Deep \& cross network for ad click predictions}. In
  \bibinfo{booktitle}{\emph{Proceedings of the ADKDD'17}}. ACM,
  \bibinfo{pages}{12}.
\newblock


\bibitem[Wang et~al\mbox{.}(2019)]%
        {wang2019neural}
\bibfield{author}{\bibinfo{person}{Xiang Wang}, \bibinfo{person}{Xiangnan He},
  \bibinfo{person}{Meng Wang}, \bibinfo{person}{Fuli Feng}, {and}
  \bibinfo{person}{Tat-Seng Chua}.} \bibinfo{year}{2019}\natexlab{}.
\newblock \showarticletitle{Neural graph collaborative filtering}. In
  \bibinfo{booktitle}{\emph{Proceedings of the 42nd international ACM SIGIR
  conference on Research and development in Information Retrieval}}.
  \bibinfo{pages}{165--174}.
\newblock


\bibitem[Wei et~al\mbox{.}(2019)]%
        {wei2019mmgcn}
\bibfield{author}{\bibinfo{person}{Yinwei Wei}, \bibinfo{person}{Xiang Wang},
  \bibinfo{person}{Liqiang Nie}, \bibinfo{person}{Xiangnan He},
  \bibinfo{person}{Richang Hong}, {and} \bibinfo{person}{Tat-Seng Chua}.}
  \bibinfo{year}{2019}\natexlab{}.
\newblock \showarticletitle{MMGCN: Multi-modal graph convolution network for
  personalized recommendation of micro-video}. In
  \bibinfo{booktitle}{\emph{Proceedings of the 27th ACM International
  Conference on Multimedia}}. \bibinfo{pages}{1437--1445}.
\newblock


\bibitem[Wu et~al\mbox{.}(2019)]%
        {wu2019session}
\bibfield{author}{\bibinfo{person}{Shu Wu}, \bibinfo{person}{Yuyuan Tang},
  \bibinfo{person}{Yanqiao Zhu}, \bibinfo{person}{Liang Wang},
  \bibinfo{person}{Xing Xie}, {and} \bibinfo{person}{Tieniu Tan}.}
  \bibinfo{year}{2019}\natexlab{}.
\newblock \showarticletitle{Session-based recommendation with graph neural
  networks}. In \bibinfo{booktitle}{\emph{Proceedings of the AAAI Conference on
  Artificial Intelligence}}, Vol.~\bibinfo{volume}{33}.
  \bibinfo{pages}{346--353}.
\newblock


\bibitem[Xin et~al\mbox{.}(2019)]%
        {xin2019cfm}
\bibfield{author}{\bibinfo{person}{Xin Xin}, \bibinfo{person}{Bo Chen},
  \bibinfo{person}{Xiangnan He}, \bibinfo{person}{Dong Wang},
  \bibinfo{person}{Yue Ding}, {and} \bibinfo{person}{Joemon Jose}.}
  \bibinfo{year}{2019}\natexlab{}.
\newblock \showarticletitle{CFM: Convolutional Factorization Machines for
  Context-Aware Recommendation.}. In \bibinfo{booktitle}{\emph{IJCAI}},
  Vol.~\bibinfo{volume}{19}. \bibinfo{pages}{3926--3932}.
\newblock


\bibitem[Xu et~al\mbox{.}(2019)]%
        {xu2019graph}
\bibfield{author}{\bibinfo{person}{Chengfeng Xu}, \bibinfo{person}{Pengpeng
  Zhao}, \bibinfo{person}{Yanchi Liu}, \bibinfo{person}{Victor~S Sheng},
  \bibinfo{person}{Jiajie Xu}, \bibinfo{person}{Fuzhen Zhuang},
  \bibinfo{person}{Junhua Fang}, {and} \bibinfo{person}{Xiaofang Zhou}.}
  \bibinfo{year}{2019}\natexlab{}.
\newblock \showarticletitle{Graph Contextualized Self-Attention Network for
  Session-based Recommendation.}. In \bibinfo{booktitle}{\emph{IJCAI}},
  Vol.~\bibinfo{volume}{19}. \bibinfo{pages}{3940--3946}.
\newblock


\bibitem[Ying et~al\mbox{.}(2018)]%
        {ying2018graph}
\bibfield{author}{\bibinfo{person}{Rex Ying}, \bibinfo{person}{Ruining He},
  \bibinfo{person}{Kaifeng Chen}, \bibinfo{person}{Pong Eksombatchai},
  \bibinfo{person}{William~L Hamilton}, {and} \bibinfo{person}{Jure Leskovec}.}
  \bibinfo{year}{2018}\natexlab{}.
\newblock \showarticletitle{Graph convolutional neural networks for web-scale
  recommender systems}. In \bibinfo{booktitle}{\emph{Proceedings of the 24th
  ACM SIGKDD International Conference on Knowledge Discovery \& Data Mining}}.
  \bibinfo{pages}{974--983}.
\newblock


\bibitem[Yu et~al\mbox{.}(2020)]%
        {yu2020tagnn}
\bibfield{author}{\bibinfo{person}{Feng Yu}, \bibinfo{person}{Yanqiao Zhu},
  \bibinfo{person}{Qiang Liu}, \bibinfo{person}{Shu Wu}, \bibinfo{person}{Liang
  Wang}, {and} \bibinfo{person}{Tieniu Tan}.} \bibinfo{year}{2020}\natexlab{}.
\newblock \showarticletitle{TAGNN: Target attentive graph neural networks for
  session-based recommendation}. In \bibinfo{booktitle}{\emph{Proceedings of
  the 43rd International ACM SIGIR Conference on Research and Development in
  Information Retrieval}}. \bibinfo{pages}{1921--1924}.
\newblock


\bibitem[Zhang et~al\mbox{.}(2017)]%
        {zhang2017joint}
\bibfield{author}{\bibinfo{person}{Yongfeng Zhang}, \bibinfo{person}{Qingyao
  Ai}, \bibinfo{person}{Xu Chen}, {and} \bibinfo{person}{W~Bruce Croft}.}
  \bibinfo{year}{2017}\natexlab{}.
\newblock \showarticletitle{Joint representation learning for top-n
  recommendation with heterogeneous information sources}. In
  \bibinfo{booktitle}{\emph{Proceedings of the 2017 ACM on Conference on
  Information and Knowledge Management}}. ACM, \bibinfo{pages}{1449--1458}.
\newblock


\bibitem[Zhou et~al\mbox{.}(2019)]%
        {zhou2019deep}
\bibfield{author}{\bibinfo{person}{Guorui Zhou}, \bibinfo{person}{Na Mou},
  \bibinfo{person}{Ying Fan}, \bibinfo{person}{Qi Pi}, \bibinfo{person}{Weijie
  Bian}, \bibinfo{person}{Chang Zhou}, \bibinfo{person}{Xiaoqiang Zhu}, {and}
  \bibinfo{person}{Kun Gai}.} \bibinfo{year}{2019}\natexlab{}.
\newblock \showarticletitle{Deep interest evolution network for click-through
  rate prediction}. In \bibinfo{booktitle}{\emph{Proceedings of the AAAI
  conference on artificial intelligence}}, Vol.~\bibinfo{volume}{33}.
  \bibinfo{pages}{5941--5948}.
\newblock


\bibitem[Zhou et~al\mbox{.}(2018)]%
        {zhou2018deep}
\bibfield{author}{\bibinfo{person}{Guorui Zhou}, \bibinfo{person}{Xiaoqiang
  Zhu}, \bibinfo{person}{Chenru Song}, \bibinfo{person}{Ying Fan},
  \bibinfo{person}{Han Zhu}, \bibinfo{person}{Xiao Ma},
  \bibinfo{person}{Yanghui Yan}, \bibinfo{person}{Junqi Jin},
  \bibinfo{person}{Han Li}, {and} \bibinfo{person}{Kun Gai}.}
  \bibinfo{year}{2018}\natexlab{}.
\newblock \showarticletitle{Deep interest network for click-through rate
  prediction}. In \bibinfo{booktitle}{\emph{Proceedings of the 24th ACM SIGKDD
  International Conference on Knowledge Discovery \& Data Mining}}.
  \bibinfo{pages}{1059--1068}.
\newblock


\end{thebibliography}

\end{document}